\documentstyle[12pt,epsf,contigrefs]{article}
\def\href#1#2{#2}   
\newif\ifdraft
\draftfalse	  
\reversemarginpar   
\makeatletter
\let\mlabel=\label
\let\adkendequation=\endequation%
\def\endequation{\adkendequation\adklabel\global\@ignoretrue}
\let\adkendeqnarray=\endeqnarray%
\def\endeqnarray{\adkendeqnarray\adklabel\global\@ignoretrue}
\newbox\marglabbox
\def\adklabel{\ifvoid\marglabbox\else\marginpar{\unhbox\marglabbox}\fi}
\def\label#1{\ifdraft\ifmmode%
  \global\setbox\marglabbox=\hbox{\hfill\fbox{\tiny\verb*~#1~}}%
  \else\ifinner\else\marginpar{\hfill\fbox{\tiny\verb*~#1~}}%
  \fi\fi\fi \mlabel{#1}}
\makeatother
\ifdraft%
\fi
\font\twelvebb=msbm12
\font\tenbb=msbm10
\font\sevenbb=msbm7
  \newfam\bbfam
  \def\bb{\fam\bbfam\twelvebb}
  \textfont\bbfam=\twelvebb
  \scriptfont\bbfam=\tenbb
  \scriptscriptfont\bbfam=\sevenbb
\font\twelveeusm=eusm10 scaled 1200
\font\teneusm=eusm10
  \newfam\eusmfam
  \def\eusm{\fam\eusmfam\twelveeusm}
  \textfont\eusmfam=\twelveeusm
  \scriptfont\eusmfam=\teneusm
  \scriptscriptfont\eusmfam=\scriptfont\eusmfam
\font\twelvefrak=eufm10 scaled 1200
\font\tenfrak=eufm10
  \newfam\frakfam
  
  \textfont\frakfam=\twelvefrak
  \scriptfont\frakfam=\tenfrak
  \scriptscriptfont\frakfam=\scriptfont\frakfam
\def\sqr#1#2{{\vcenter{\hrule height.#2pt
   \hbox{\vrule width.#2pt height#1pt \kern#1pt
      \vrule width.#2pt}
   \hrule height.#2pt}}}

\def\bsqr#1#2{{\vrule width #1pt height#2pt}}
\def\bsquare{{\mathchoice\bsqr66\bsqr66\bsqr33\bsqr33}}
\def\badbreak{\penalty1000}
\newtheorem{theorem}{Theorem}
\newtheorem{lemma}{Lemma}
\newtheorem{definition}{Definition}
\newtheorem{corollary}{Corollary}
\newenvironment{proof}{{\em Proof.}}{\badbreak$\;\bsquare$\smallskip}
\def\Adj{\mathop{\rm Adj}}		    
\def\deg{\mathop{\rm deg}}		    
\def\var{\mathop{\rm Var}}		    
\def\identity{{\bb I}}			    
\def\implies{\Rightarrow}		    

\def\union{\cup}
\def\intersection{\cap}
\def\rational#1#2{{\mathchoice{\textstyle{#1\over#2}}%
  {\scriptstyle{#1\over#2}}{\scriptscriptstyle{#1\over#2}}{#1/#2}}}
\def\half{\rational12}			    
\def\R{{\bb R}}				    
\def\Z{{\bb Z}}				    
\def\C{{\bb C}}				    
\newcommand{\zint}{{z_{\mbox{\tiny int}}}}  
\newcommand{\zexp}{{z_{\mbox{\tiny exp}}}}  
\newcommand{\hmd}{Hybrid Molecular Dynamics}  
\newcommand{\lhmc}{Local Hybrid Monte Carlo}  
\newcommand{\phim}{{\tilde\phi}}	    
\newcommand{\freq}{f}			    
\newcommand{\freqeqm}{\omega}		    
\newcommand{\M}{{\cal M}}		    
\newcommand{\Msq}{{{\cal M}^2}}	            
\newcommand{\F}{{\cal F}}		    
\newcommand{\orp}{\zeta}		    
\newcommand{\A}{{\cal A}}		    
\newcommand{\ace}{\tau}			    
\newcommand{\acf}{{\cal C}^c}		    
\newcommand{\cacf}{{\eusm C}^c}		    
\def\emu#1{\hat{#1}}	  		    
\newcommand{\pcen}{{p_c}}		    
\newcommand{\ppcen}{{p-\pcen}}		    
\newcommand{\bppl}{{\beta^+_p}}	            
\newcommand{\bpmi}{{\beta^-_p}}	            
\newcommand{\bmi}{{\beta^-}}	            
\newcommand{\bpl}{{\beta^+}}	            

\newcommand{\PB}{\alpha^{-p}}
\newcommand{\QQ}{\alpha^q}
\newcommand{\QB}{\alpha^{-q}}

\newcommand{\PBM}{\alpha^{-p_\mu}}
\newcommand{\QQM}{\alpha^{q_\mu}}

\newcommand{\PBN}{\alpha^{-p_\nu}}

\newcommand{\QBE}{\alpha^{-q_\epsilon}}

\begin{document}

\title{
  \vskip-1in\smash{\small\vbox{\rightline{FSU--SCRI--97--94}
    \rightline{hep--lat/9708024}}}\vskip0.7in
  The {\lhmc} Algorithm for\\ Free Field Theory:\\
  Reexamining Overrelaxation}

\author{Ivan Horv\'ath\thanks{{\tt horvath@scri.fsu.edu}}~~and
    A.~D.~Kennedy\thanks{{\tt adk@scri.fsu.edu}} \\[1ex]
  Supercomputer Computations Research Institute, \\
  Florida State University, Tallahassee, Florida 32306-4052, U.S.A.}
\date{\today}

\maketitle

\begin{abstract}
  \noindent We analyze the autocorrelations for the {\lhmc} algorithm
  \cite{kennedy92c} in the context of free field theory. In this case this is
  just Adler's overrelaxation algorithm \cite{adler81a}. We consider the
  algorithm with even/odd, lexicographic, and random updates, and show that its
  efficiency depends crucially on this ordering of sites when optimized for a
  given class of operators. In particular, we show that, contrary to previous
  expectations, it is possible to eliminate critical slowing down ($\zint=0$)
  for a class of interesting observables, including the magnetic susceptibility:
  this can be done with lexicographic updates but is not  possible with even/odd
  ($\zint=1$) or random ($\zint=2$) updates. We are considering the dynamical
  critical exponent $\zint$ for integrated autocorrelations rather than for the
  exponential autocorrelation time; this is reasonable because it is the
  integrated autocorrelation which determines the cost of a Monte Carlo
  computation.\parfillskip=0pt\par
\end{abstract}

\section{Introduction}

Stochastic overrelaxation, and especially its variant usually referred to as
``hybrid overrelaxation,'' is generally considered to be the most efficient
algorithm for generating lattice configurations in pure gauge theory. As such it
is also frequently used for the purposes of quenched simulations, although at
the currently studied (small) correlation lengths the relative improvement over
the standard Metropolis or heatbath local algorithms is a priori rather modest.

The idea of generalizing overrelaxation methods to the stochastic case is due to
Adler~\cite{adler81a}, who proposed an algorithm for multiquadratic actions such
as free field theory (see also~\cite{whitmer84a}). The initial expectations of
improved performance were confirmed by studying the dynamical critical behaviour
of the algorithm in this context
analytically~\cite{adler88a,adler88b,neuberger87a,wolff92a}. The main result of
these studies was that the algorithm can be tuned to achieve the dynamical
critical exponent $\zexp=1$ as compared to the generic value $\zexp=2$ for the
standard local Metropolis or heatbath algorithms. Some additional theoretical
insights were also obtained in references~\cite{sokal89a,neuberger92a}.

Several extensions of the algorithm to interacting field theories and subsequent
partial modifications were introduced by various authors,
including~\cite{adler88b,brown87a,creutz87c,heller89b,petronzio90a,kennedy93a}.
These algorithms were studied in more detail in several numerical works such
as~\cite{gupta88e,adler89a,baillie91b,akemi93b}. While these almost invariably
claim useful improvement, the truly systematic study (such that could make
reasonable conclusions about dynamical critical exponents) is still missing.

The purpose of this paper is to analyze the free field behaviour of
overrelaxation further. Apart from our original aim of evaluating certain
autocorrelations, we obtained some interesting conceptual insights which we feel
are worth communicating. It has become clear over the years that the performance
of Monte Carlo algorithm should be assessed with respect to a given operator
$\Omega$ (or set of operators): in particular, the most relevant characteristic
then is the dynamical critical exponent $z_\Omega$, corresponding to the
integrated autocorrelation for that quantity. While the dynamical critical
exponent for the exponential autocorrelation time (which we denoted by $\zexp$
in previous paragraphs) usually gives us an upper bound, it often happens that
$z_\Omega<\zexp$ for most observables.

If we have a situation where for operators of interest, $z_\Omega$ can be made
smaller than $\zexp$, then the latter is somewhat irrelevant for the task at
hand. In lattice field theory we actually have a specific group of operators we
want to speed up, namely the ones for which the low energy (momentum) states are
most important, because these operators are relevant for the continuum limit. We
should thus attempt to optimize the overrelaxation algorithm for these
quantities, rather than minimize $\zexp$. It has been implicitly assumed in
previous studies that zero momentum operators (such as the magnetic
susceptibility) actually saturate the $\zexp$-bound and so the above distinction
is just a wishful thinking. Our main point in this paper is that while this is
true for even/odd updates, where the dynamics of overrelaxation intrinsically
couples the low and high frequency modes, it is not true for lexicographic
updates, where the modes decouple at large volumes. In fact in the latter case
we can tune the overrelaxation algorithm such that critical slowing down is
completely eliminated for quantities that depend only on the zero momentum mode.
We emphasize that this is a different tuning than the one leading to $\zexp=1$;
in fact, with this tuning $\zexp=2$. However, in this case the $\zexp$-bound is
actually saturated by high frequency quantities most of which we are not
interested in.

In the light of the above discussion, the ``conventional wisdom'' about the
inability of local algorithms to perform better than $\zexp\approx 1$ (see for
example~\cite{neuberger92a}) can be rather misleading. While the statement as
such might be true (but has not been proved even for free field theory), it does
not tell us much about how the local algorithm will perform in the specific case
of interest.

The second point we want to make concerns the introduction of the additional
noise in the overrelaxation update. In particular we attempted the same
optimization for a scheme in which the updated site is chosen at random from a
uniform distribution. It turns out that in this case it is impossible to tune
the overrelaxation parameter to reduce the critical slowing down for zero
momentum quantities; in fact we cannot do better than $z_\Omega=2$ for these
operators. This means that overrelaxation can easily lose its magic if we
introduce extra noise in the procedure.

In~\cite{kennedy92c} it was shown that the overrelaxation algorithm for free
field theory is a special case of the class of {\hmd} algorithms where the
fields are updated one site at a time using an analytic solution of the
equations of motion. This rather suprising and elegant connection is especially
intriguing in the light of the fact that, as shown in~\cite{kennedy93a}, a
corresponding algorithm can be found for gauge theories. Here we adopt this
point of view and will frequently refer to free field overrelaxation as
{\lhmc}~(LHMC).

We start in section~\ref{sec:lhmc-fft} by formulating the LHMC algorithm and
developing the techniques to calculate the integrated autocorrelations for
typical zero momentum quantities, namely magnetization and susceptibility. In
section~\ref{sec:updating-schemes} we analyze the performance of LHMC with
even/odd, lexicographic, and random updates. We also give a proof that the
integrated autocorrelation for the magnetic susceptibility vanishes at large
volume with optimally tuned overrelaxation parameter when using the modified
lexicographic update. In Appendix~\ref{sec:appendix-A} we give an explicit
asymptotic calculation of the update matrix for the lexicographic scheme, and in
Appendix~\ref{sec:appendix-B} we develop a formalism for bounding the finite
volume corrections using methods of functional analysis.

\section{{\lhmc} for Free Field Theory} \label{sec:lhmc-fft}

\subsection{Analytic Solution}

We wish to consider a real scalar free field theory described by a functional
integral with the usual action
\begin{displaymath}
  S[\phi] \equiv \half\sum_{x\in\Z_L^d} \{-\phi_x\Delta\phi_x + m^2\phi_x^2\},
\end{displaymath}
and a flat (Lebesgue) measure for the $\phi$ field. The lattice Laplacian
operator is defined as
\begin{displaymath}
  \Delta\phi_x \equiv
    \sum_{\mu=1}^d (\phi_{x+\emu\mu} - 2\phi_x + \phi_{x-\emu\mu}),
\end{displaymath}
where $\emu\mu$ is a unit vector in the $\mu$ direction. Since the theory is
free it can be diagonalized using the Fourier transformed fields
\begin{displaymath}
  \phi_x \equiv L^{-d/2} \sum_{p\in\Z_L^d} \phim_p e^{2\pi ip\cdot x/L},
\end{displaymath}
where $\phim_p^* = \phim_{-p}$ since $\phi_x\in\R$.\footnote{For notational
convenience we will frequently omit the tildes and follow the convention that
Fourier components are implied by subscripts $p,q,\ldots$ as opposed to
$x,y,\ldots$.}
The action simplifies to
\begin{equation}
  S[\phim] = \half\sum_{p\in\Z_L^d} \freq_p^2 |\phim_p|^2
  \label{eq:d10}
\end{equation}
with
\begin{displaymath}
  \freq_p^2 \equiv m^2 + 2d - 2\sum_{\mu=1}^d \cos{2\pi p_\mu\over L} \;.
\end{displaymath}
In particular we note that the lowest and highest frequencies of the system are
$\freq_{(0,\ldots,0)}^2 = m^2$ and $\freq_{(\half L,\ldots,\half L)}^2 = m^2 +
4d$.

\subsection{{\lhmc} Updates}

Consider the dependence of the action $S$ upon one degree of freedom only,
$S(\phi_x) = \half\freqeqm^2\phi_x^2 - \F_x\phi_x + \mbox{constant}$, where here
$\freqeqm^2 \equiv m^2 + 2d$ and\footnote{It is easy to get a factor of two
wrong here!} $\F_x \equiv \sum_{\mu=1}^d (\phi_{x+\emu\mu} + \phi_{x-\emu\mu})$.
Introducing the fictitious Hamiltonian $H(\phi_x, \pi_x) \equiv \half\pi_x^2 +
S(\phi_x)$ we have
\begin{displaymath}
  H = \half\left[ \pi_x^2 + \freqeqm^2\left(\phi_x -
    {\F_x\over\freqeqm^2}\right)^2 \right] + \mbox{constant},
\end{displaymath}
and the solution of the corresponding equations of motion is
\begin{displaymath}
  \phi_x(t) - {\F_x\over\freqeqm^2} =
    \left(\phi_x(0) - {\F_x\over\freqeqm^2}\right) \cos \freqeqm t
    + {\pi_x(0)\over\freqeqm} \sin \freqeqm t,
\end{displaymath}
or
\begin{equation}
  \phi'_x = (1-\orp)\phi_x + {\sqrt{\orp(2-\orp)}\over\freqeqm} \pi_x
    + {\F_x\over\freqeqm^2} \orp
  \label{eq:d20}
\end{equation}
in terms of the {\em overrelaxation parameter\/} $\orp \equiv 1 - \cos\freqeqm
t$.

Assembling the field variables and their conjugate fictitious momenta into the
vectors
\begin{equation}
  \phi\equiv\left(
    \begin{array}{c}
      \phi_{x_1} \\ \phi_{x_2} \\ \vdots \\ \phi_{x_{L^d}}
    \end{array}\right)\qquad\mbox{and}\qquad \pi\equiv\left(
    \begin{array}{c}
      \pi_{x_1} \\ \pi_{x_2} \\ \vdots \\ \pi_{x_{L^d}}
    \end{array}\right)
  \label{eq:d25}
\end{equation}
we can write the elementary local update (\ref{eq:d20}) in the form
\begin{equation}
  \phi \mapsto M_x\phi + P_x\pi;
  \label{eq:d30}
\end{equation}
here $M_x$ and $P_x$ are $L^d\times L^d$ matrices, and the subscript refers to
the site being updated. More explicitly, we have
\begin{eqnarray}
  (M_z)_{x,y} &=& \delta_{x,y} + \delta_{x,z}\Bigl[-\orp\delta_{x,y} +
    {\orp\over\freqeqm^2} \sum_{\mu=1}^d
      (\delta_{x+\emu\mu,y} + \delta_{x-\emu\mu,y})\Bigr],
    \label{eq:d35} \\
  (P_z)_{x,y} &=& \delta_{x,y}\delta_{x,z} {\sqrt{\orp(2-\orp)}\over\freqeqm}.
    \nonumber
\end{eqnarray}
Definition (\ref{eq:d25}) implicitly assumes that we have ordered the field
variables in a certain way. Sweeping through the lattice in this order, the
complete update is given by
\begin{eqnarray}
  \phi &\mapsto& M_{x_1}\phi + P_{x_1}\pi
    \mapsto (M_{x_2}M_{x_1})\phi + (M_{x_2}P_{x_1}+P_{x_2})\pi \nonumber \\
    \qquad && \mapsto \cdots \mapsto \left(\prod_{j=1}^{L^d} M_{x_j}\right)\phi
      + \left[\sum_{i=1}^{L^d}
	\left(\prod_{j=i+1}^{L^d} M_{x_j}\right) P_{x_i}\right]\pi
    \equiv M\phi + P\pi.
  \label{eq:d40}
\end{eqnarray}
Note that the form of the matrices $M$ and $P$ depends upon the predetermined
order in which we chose to sweep through the lattice. In fact, as we shall
discuss, the efficiency of the algorithm (optimized for a given operator) can
depend crucially on this order.

In addition to such fixed-order updates we will also analyze the algorithm in
which the updated sites are chosen at random. The formul{\ae}
(\ref{eq:d30}--\ref{eq:d40}) carry over unchanged to this case except that the
updated site becomes an uniformly distributed random variable over~$\Z_L^d$.

\subsection{Cost of Measuring the Operators}

If we are interested in measuring the expectation value of some operator
$\Omega$, and $\bar\Omega$ is the average over a sequence of $T$ configurations
generated by some Markov process, we expect
\begin{displaymath}
  \bar\Omega = \langle\Omega\rangle
    \pm \sqrt{{2\A_\Omega + 1\over T}\var\Omega}.
\end{displaymath}
Here $\var\Omega = \langle\Omega^2\rangle - \langle\Omega\rangle^2$ is the
intrinsic variance of $\Omega$, and $\A_\Omega$ is the {\em integrated
autocorrelation} defined as
\begin{displaymath}
  \A_\Omega \equiv \sum_{t=1}^\infty
    {\Bigl\langle\Omega[\phi(t)]\Omega[\phi(0)]\Bigr\rangle -
     \Bigl\langle\Omega[\phi(t)]\Bigr\rangle\Bigl\langle\Omega[\phi(0)]
     \Bigr\rangle
  \over
     \bigl\langle\Omega^2\bigr\rangle -
     \bigl\langle\Omega\bigr\rangle^2} \equiv
     \sum_{t=1}^\infty \acf_\Omega(t),
\end{displaymath}
where $\acf_\Omega$ is the connected autocorrelation function. The cost of
measuring $\langle\Omega\rangle$ to a given accuracy depends upon $\A_\Omega$,
and for local algorithms (such as LHMC) this is the only relevant characteristic
which we need to compute in order to ascertain the performance of the algorithm.
For a large system near criticality
\begin{displaymath}
   \A_\Omega \sim \xi^{z_\Omega} \qquad (1\ll\xi\ll L),
\end{displaymath}
where $z_\Omega$ is the dynamical critical exponent, corresponding to quantity
$\Omega$.\footnote{Note that our definition of the dynamical critical exponent
assumes that the large volume limit is taken first.}

In this paper we are interested in studying the performance of LHMC when applied
to measuring some interesting operators in the context of free field theory. To
this end we shall mainly consider the magnetization $\M$ and magentization
squared $\Msq$, where we define the magnetization as
\begin{displaymath}
  \M \equiv {1\over L^{d/2}}\sum_{x\in\Z_L^d} \phi_x = \phim_0 .
\end{displaymath}
The expectation values of powers of $\M$ are given by
\begin{displaymath}
  \langle\M^\alpha\rangle = \langle\phim_0^\alpha\rangle =
    {\int_{-\infty}^\infty d\phim_0\,
      e^{-\half\freq_0^2\phim_0^2} \phim_0^\alpha \over
    \int_{-\infty}^\infty d\phim_0\,e^{-\half\freq_0^2\phim_0^2}}
  = \left\{
    \begin{array}{ll}
      0 & \mbox{if $\half\alpha\not\in\Z$,} \\
      {\Gamma\left({\alpha+1\over2}\right) \over \Gamma(\half)}
	\left(2\over\freq_0^2\right)^{\alpha/2}
	  & \mbox{if $\half\alpha\in\Z$.}
    \end{array}\right.
\end{displaymath}
In particular, $\langle\M\rangle=0$, $\langle\M^2\rangle={1\over m^2}$ and
$\langle\M^4\rangle={3\over m^4}$. We thus expect that the measured values will
satisfy $\overline\M = 0\pm {1\over m}\sqrt{2\A_\M + 1\over T}$, and
$\overline{\Msq} = {1\over m^2} \pm {\sqrt{2}\over m^2}\sqrt{2\A_\Msq+1 \over
T}$.

\subsection{Autocorrelations}

The general formalism for calculating the autocorrelations of any operator that
is polynomial in the fields has been developed in \cite{kennedy91a}. Here we
shall use the main ideas of that approach and derive explicit formul{\ae} for
operators we are interested in.

For convenience we will work in momentum space. In particular, we assume that
the momentum-space representation of matrices $M$ and $P$ that characterize the
updating scheme of our choice is known (see equation (\ref{eq:d40})) and we
express the autocorrelations in terms of the corresponding matrix elements.

\subsubsection{Magnetization}

If we are interested in the Monte Carlo evolution of quantities linear in field
variables, we can average the relation~(\ref{eq:d40}) over the independent
fictitious momenta, and write
\begin{displaymath}
  \langle\phi_p\rangle_\pi \equiv \langle\phi_p(0)\rangle_\pi
    \mapsto \langle\phi_p(1)\rangle_\pi = M_{p,q}\langle\phi_q\rangle_\pi.
\end{displaymath}
Consequently we have
\begin{equation}
  1 + \A_\M = \sum_{t=0}^\infty \acf_\M(t)
   = \sum_{t=0}^\infty {\left\langle\phi_0(M^t)_{0,q}\phi_q\right\rangle\over
      \langle\phi_o^2\rangle}
   = \sum_{t=0}^\infty (M^t)_{0,0}
   = (\identity - M)^{-1}_{0,0}\;,
  \label{eq:d60}
\end{equation}
where we have used the relation
\begin{displaymath}
  \langle\phi_p\phi_q\rangle = {\delta_{p+q,0}\over\freq_p^2}.
\end{displaymath}

An explicit representation for the normalized connected autocorrelation function
$\acf_\M(t)$ may be obtained by introducing an auxilliary variable $x$ and
evaluating the generating function
\begin{displaymath}
  1 + \A_\M(x) \equiv \sum_{t=0}^\infty x^t\acf_\M(t)
    = \sum_{t=0}^\infty (xM)^t_{0,0} = (\identity - xM)^{-1}_{0,0}.
\end{displaymath}
On the right hand side we have a rational expression $Q(x)/R(x)$, where the
degrees of polynomials satisfy the conditions\footnote{If the matrix
$(\identity-xM)^{-1}$ is block diagonal then $\deg R$ is equal to the dimension
of the block containing the element $(\identity-xM)^{-1}_{0,0}$.}
\begin{displaymath}
  \deg R \leq L^D, \qquad \deg Q < \deg R.
\end{displaymath}
We can thus perform the partial fraction expansion of $Q(x)/R(x)$ and
write\footnote{The generalization to the case where $R$ has degenerate roots is
trivial.}
\begin{displaymath}
   \sum_{t=0}^\infty x^t\acf_\M(t)
    = \sum_{j=1}^{\deg R} {a_j\over 1-b_j x}
    = \sum_{j=1}^{\deg R} a_j \sum_{t=0}^\infty \left(b_j x\right)^t.
\end{displaymath}
Equating the coefficients of $x^t$ we can write the autocorrelation function in
terms of the coefficients in the partial fraction expansion, namely
\begin{displaymath}
  \acf_\M(t) = \sum_{j=1}^{\deg R} a_j e^{t\ln b_j}.
\end{displaymath}
Note that this also allows us to express the cumulative autocorrelation function
in the form
\begin{displaymath}
  \cacf_\M(T) \equiv \sum_{t=0}^T \acf_\M(t)
    = \sum_{j=1}^{\deg R} a_j {1 - b_j^{(T+1)}\over 1 - b_j}
\end{displaymath}
and the integrated autocorrelation as
\begin{displaymath}
  1 + \A_\M = \lim_{t\to\infty} \cacf_\M(T)
    = \sum_{j=1}^{\deg R} {a_j\over1 - b_j} = {Q(1)\over R(1)}.
\end{displaymath}

\subsubsection{Quadratic operators}

Turning now to quadratic operators we consider a generic quantity of the form
\begin{equation}
  \Omega = \sum_q \phi^*_q\phi_q K_q
  \label{eq:d81}
\end{equation}
with the spectral density $K_q$ being some function of the momentum. The change
of the quadratic monomial $\phi^*_p\phi_q$ after a single sweep is given by
\begin{displaymath}
  \phi^*_p\phi_q \mapsto {\phi'}^*_p{\phi'}_q
    = \sum_{r,s} (M_{p,r}\phi_r + P_{p,r}\pi_r)^*
      (M_{q,s}\phi_s + P_{q,s}\pi_s),
\end{displaymath}
which can be averaged over fictitious momenta to give
\begin{displaymath}
  \left\langle{\phi'}^*_p{\phi'}_q\right\rangle_\pi
    \mapsto \sum_{r,s} M_{p,r}^*M_{q,s}
      \left\langle\phi^*_r\phi_s\right\rangle_\pi + \sum_r P_{p,r}^* P_{q,r}.
\end{displaymath}
We can express this as
\begin{equation}
  \langle{\phi^Q}'\rangle_\pi = M^Q\langle\phi^Q\rangle_\pi + P^Q,
  \label{eq:av-quad-update}
\end{equation}
where
\begin{displaymath}
  \phi^Q_{p,q} \equiv \phi^*_p\phi_q,
  \qquad M^Q_{pq,rs} \equiv (M^*\otimes M)_{pq,rs} = M^*_{p,r} M_{q,s},
  \qquad \mbox{and} \quad P^Q_{p,q} \equiv \sum_r P^*_{p,r}P_{q,r}.
\end{displaymath}
The integrated autocorrelation is then given by
\begin{displaymath}
  1 + \A_{\Omega} =
      \sum_{t=0}^\infty {\sum_{q_1,q_2,r,s} K_{q_1}K_{q_2} [(M^Q)^t]_{q_2q_2,rs}
      \bigl[\langle\phi^*_{q_1}\phi_{q_1}\phi^*_r\phi_s\rangle
	- \langle\phi^*_{q_1}\phi_{q_1}\rangle\langle\phi^*_r\phi_s\rangle\bigr]
      \over \sum_{q_1,q_2} K_{q_1}K_{q_2}
	\bigl[\langle\phi^*_{q_1}\phi_{q_1}\phi^*_{q_2}\phi_{q_2}\rangle
	- \langle\phi^*_{q_1}\phi_{q_1}\rangle
	  \langle\phi^*_{q_2}\phi_{q_2}\rangle\bigr]}.
\end{displaymath}
Performing the sum over $t$ and using the relation
\begin{displaymath}
  \langle\phi^*_q\phi_q\phi^*_r\phi_s\rangle
    - \langle\phi^*_q\phi_q\rangle \langle\phi^*_r\phi_s\rangle
      = 2\delta_{q,r}\delta_{q,s} \freq_q^{-4},
\end{displaymath}
we find
\begin{equation}
  1 + \A_{\Omega} =
      {\sum_q K_q^2 (\identity - M^Q)^{-1}_{qq,qq} \freq^{-4}_q
       \over \sum_q K_q^2 \freq^{-4}_q}.
  \label{eq:d100}
\end{equation}
Note that the matrix elements of the identity matrix in the quadratic basis are
$\identity_{pq,rs}\equiv \delta_{p,r}\delta_{q,s}$.

In the case of $\Msq$ we have $K_q=\delta_{q,0}$, which implies
\begin{displaymath}
  1 + \A_\Msq = [\identity -M^Q]^{-1}_{00,00}.
\end{displaymath}
Note the formal similarity of this expression to that of equation~(\ref{eq:d60})
for the magnetization.

In what follows we will also refer to the integrated autocorrelation for the
energy (action)~$E$; in this case we have $K_q=\half\freq_q^2$, giving
\begin{displaymath}
  1 + \A_{E} = {1\over L^d} \sum_q [\identity-M^Q]^{-1}_{qq,qq}.
\end{displaymath}

There is obviously a certain redundancy in the matrix $M^Q$, since the basis
elements we have used are not independent. In particular $\phi^*_p\phi_q =
(\phi^*_q\phi_p)^*$. In practical calculations it is usually advantageous to
reduce the basis to its independent subset, thus reducing the dimensionality of
the matrices involved. Also, in case of $\M^2$ we can use the basis $\{\phi_p
\phi_q\}$ instead of $\{\phi^*_p \phi_q\}$.

Notice also that the formalism for calculating autocorrelation functions which
we described in detail for the magentization carries over unchanged; calculating
the $\acf_{\Omega}(t)$ involves the partial fraction expansion of
$\A_{\Omega}(x)$.

\section{Three Updating Schemes} \label{sec:updating-schemes}

The exact analysis of autocorrelations for the linear stochastic update
(\ref{eq:d40}) is rather intractable in the general case, i.e., for an arbitrary
ordering of the updated sites. However, in special cases the equations simplify
and an exact analysis can be performed. In what follows, we will analyze the
even/odd, lexicographic, and random updates.

From our discussion in previous sections it follows that our work will split
into two steps: First we need to find the Fourier representation of matrices $M$
and $P$ corresponding to the updating scheme in question, and then we must
evaluate the relevant formul{\ae} for autocorrelations. Note that the explicit
form of matrix $P$ is not needed for quantities we are interested in; this,
however, is not true in general.

\subsection{Even/Odd Updates}

Consider splitting the lattice into odd and even sites, as on a checkerboard. If
we choose to update all the sites of a given sublattice before the sites of the
other sublattice, then the result does not depend on the order of sites within
the sublattices. Indeed, updating all the even (odd) sites we may write
\begin{displaymath}
  \phi'_x = \phi_x + \chi_\pm(x) \left\{-\orp\phi_x +
    {\sqrt{\orp(2-\orp)}\over\freqeqm} \pi_x + \orp {\F_x\over\freqeqm^2}
      \right\},
\end{displaymath}
where $\chi_\pm(x) \equiv \half\left[1\pm(-1)^{\sum_{\mu=1}^d x_\mu}\right]$
which is one for even (odd) sites and is zero otherwise.

The function $\chi_\pm(x)$ has a simple Fourier representation,
\begin{displaymath}
  \chi_\pm(x) = \half \left[e^{2\pi i 0\cdot x \over L} \pm
                            e^{2\pi i \pcen\cdot x\over L}\right],
\end{displaymath}
where $\pcen\equiv\half(L,\ldots,L)$, so we obtain in Fourier
space\footnote{Note that $\cos{2\pi(\ppcen)_\mu\over L} = -\cos{2\pi p_\mu\over
L}$.}
\begin{eqnarray*}
  \phi'_p &=& \phi_p + L^{-d/2} \sum_{x\in\Z_L^d} e^{-2\pi i p\cdot x/L}
    \chi_\pm(x) \left\{ - \orp\phi_x + {\sqrt{\orp(2-\orp)}\over\freqeqm} \pi_x
    + \orp {\F_x\over\freqeqm^2} \right\} \\
  &=& \phi_p + \half \Biggl\{ -\orp(\phi_p \pm \phi_\ppcen)
    + {\sqrt{\orp(2-\orp)}\over\freqeqm} (\pi_p \pm \pi_\ppcen) \\
  &&\qquad\qquad + {2\orp\over\freqeqm^2}
    \sum_{\mu=1}^d \left(\cos{2\pi p_\mu\over L}\phi_p
      \mp \cos{2\pi p_\mu\over L}\phi_\ppcen \right) \Biggr\}.
\end{eqnarray*}
In matrix notation we have
\begin{displaymath}
  \left(
    \begin{array}{c}
      \phi_p \\
      \phi_\ppcen
    \end{array}
  \right)' = \left(
    \begin{array}{cc}
      1 + \bppl & \pm \bpmi \\
      \pm \bppl & 1 + \bpmi
    \end{array}
  \right) \left(
    \begin{array}{c}
      \phi_p \\
      \phi_\ppcen
    \end{array}
  \right) + {\sqrt{\orp(2-\orp)}\over2\freqeqm} \left(
    \begin{array}{cc}
      1 & \pm1 \\
      \pm1 & 1
    \end{array}
  \right) \left(
    \begin{array}{c}
      \pi_p \\
      \pi_\ppcen
    \end{array}
  \right),
\end{displaymath}
where $\beta^\pm_p \equiv \half\orp\left(-1\pm{2\over\freqeqm^2}
\sum_{\mu=1}^d \cos {2\pi p_\mu\over L}\right)$.

For an even update followed by an odd update, which is the fundamental ergodic
Markov step, we have\footnote{If we are careful we observe that the equations of
motion for the fictitious momenta are
\begin{displaymath}
  \pi'_x = \pi_x + \chi_\pm(x) \left\{ - \orp\pi_x +
    {\sqrt{\orp(2-\orp)}\over\freqeqm} (-\freqeqm^2\phi_x + \F) \right\}
\end{displaymath}
or
\begin{displaymath}
  \left(
    \begin{array}{c}
      \pi_p \\
      \pi_\ppcen
    \end{array}
  \right)' = \left(
    \begin{array}{cc}
      1 - \half\orp & \mp \half\orp \\
      \mp \half\orp & 1 - \half\orp
    \end{array}
  \right) \left(
    \begin{array}{c}
      \pi_p \\
      \pi_\ppcen
    \end{array}
  \right) + \freqeqm\sqrt{\orp(2-\orp)} \left(
    \begin{array}{cc}
      \bppl & \pm \bpmi \\
      \pm \bppl & \bpmi
    \end{array}
  \right) \left(
    \begin{array}{c}
      \phi_p \\
      \phi_\ppcen
    \end{array}
  \right),
\end{displaymath}
but
\begin{displaymath}
  \left(
    \begin{array}{cc}
      1 & \mp1 \\
      \mp1 & 1
    \end{array}
  \right) \left(
    \begin{array}{c}
      \pi_p \\
      \pi_\ppcen
    \end{array}
  \right)' = \left(
    \begin{array}{cc}
      1 & \mp1 \\
      \mp1 & 1
    \end{array}
  \right) \left(
    \begin{array}{c}
      \pi_p \\
      \pi_\ppcen
    \end{array}
  \right),
\end{displaymath}
so it makes no difference whether we evolve the momenta or not. Of course, this
just reflects the fact that the momenta on even and odd sites are independent
random variables.}
\begin{displaymath}
  \left(
    \begin{array}{c}
      \phi_p \\
      \phi_\ppcen
    \end{array}
  \right) \;\mapsto\; M_p \left(
    \begin{array}{c}
      \phi_p \\
      \phi_\ppcen
    \end{array}
  \right) + P_p \left(
    \begin{array}{c}
      \pi_p \\
      \pi_\ppcen
    \end{array}
  \right),
\end{displaymath}
with
\begin{displaymath}
  M_p \equiv \left(
    \begin{array}{cc}
      1 + \bppl(2+\bppl-\bpmi) & \bpmi(\bppl-\bpmi) \\
      \bppl(\bpmi-\bppl) & 1 + \bpmi(2+\bpmi-\bppl)
    \end{array}
  \right),
\end{displaymath}
and
\begin{displaymath}
  P_p \equiv {\sqrt{\orp(2-\orp)}\over2\freqeqm} \left(
    \begin{array}{cc}
      2 + \bppl - \bpmi & \bppl - \bpmi \\
      \bpmi - \bppl & 2 + \bpmi - \bppl
    \end{array}
  \right).
\end{displaymath}

Note that with even/odd updates the Fourier modes of the field are only coupled
in pairs. In particular, the lowest energy mode $\phi_0$ is only coupled to the
highest energy mode $\phi_\pcen$; for this case we will abbreviate
$\beta^\pm_0\equiv\beta^\pm=\half\orp (-1\pm 2d/\freqeqm^2)$.

\subsubsection{Autocorrelations for $\M$}

Using the formula~(\ref{eq:d60}) we can now trivially calculate the integrated
autocorrelation for $\M$, namely
\begin{equation}
  \A_\M = (\identity-M_0)^{-1}_{0,0} - 1
    = {\bpl-\bmi-2\over 4\bpl} - 1
    = {1\over\orp}\left(1+{2d\over m^2}\right) - \left(1+{d\over m^2}\right).
  \label{eq:eoac}
\end{equation}
Since $\orp\in[0,2]$, this attains its minimum (as a function of $\orp$) for
$\orp=2$, where $\A_\M = -\half$, giving $\overline\M=0\pm0$. This is of course
true even though, or rather because, $\orp=2$ does not correspond to an ergodic
algorithm. If we tune the overrelaxation parameter  to $\orp=\orp^c\equiv
(m^2+2d)/(m^2+d)$ we have $\A_\M\equiv 0$, implying $z_\M=0$ while the algorithm
is ergodic. It should be emphasized though that this does not necessarily mean
that the algorithm generates an independent estimate of $\M$ after every sweep:
indeed, let us calculate the autocorrelation function $\acf_\M(t)$. According to
our general discussion of autocorrelations, this requires the partial fraction
decomposition of
\begin{displaymath}
   (\identity-xM_0)^{-1}_{0,0} = {Q(x)\over R(x)}
     = {Q_0 + Q_1 x \over R_0 + R_1 x + R_2 x^2}
     = \sum_{j=1}^2 {a_j\over 1-b_j x};
\end{displaymath}
in particular, we have
\begin{equation}
  \begin{array}{lll}
    Q_0 = 1, & \qquad Q_1 = -(\bmi + 1)^2 + \bpl\bmi, & \\
    R_0 = 1, & \qquad R_1 = -2(\bpl + \bmi + 1) - (\bpl - \bmi)^2, & \qquad
    R_2 = (\bpl + \bmi + 1)^2.
  \end{array}
\end{equation}
If $x_1$ and $x_2$ are the roots of the equation $R(x)=0$, then
\begin{equation}
  a_1 = {Q_0 + Q_1 x_1 \over R_2(x_2-x_1)x_1}, \qquad
  a_2 = {Q_0 + Q_1 x_2 \over R_2(x_1-x_2)x_2}, \qquad
  b_1 = {1\over x_1}, \qquad b_2 = {1\over x_2}.
  \label{eq:eo-mag-acfcoeff}
\end{equation}
Note that since $x_2=x_1^*$, we have $a_2=a_1^*$ and $b_2=b_1^*$ as one would
expect since the autocorrelation function $\acf_\M(t)=a_1b_1^t + a_2b_2^t$ is
real.

For $\orp=\orp^c$ we have in particular
\begin{displaymath}
  a_1=\half + i{m\over 4\sqrt{d}}, \qquad
  a_2 = a_1^*, \qquad
  b_1 = -{d\over(\sqrt d-im)^2}, \qquad
  b_2 = b_1^*;
\end{displaymath}
thus the autocorrelation function has an oscillatory behaviour with regions of
correlation and anticorrelation. It is only the integrated autocorrelation that
sums to zero for any $m$, giving $z_\M=0$.

\subsubsection{Autocorrelations for $\Msq$}

The update matrix $M$ is block-diagonal, so it suffices to consider  the
$4\times4$ block $M^Q_p$ spanned by the basis $\{\phi_p^*\phi_p,
\phi_p^*\phi_\ppcen, \phi_\ppcen^*\phi_p, \phi_\ppcen^*\phi_\ppcen\}$. The
matrix elements of $M$ and the above basis are real so this can be further
reduced, and we use
\begin{displaymath}
  \phi^Q_p \equiv \left(
    \begin{array}{c}
      \phi_p^*\phi_p \\
      \phi_p^*\phi_\ppcen \\
      \phi_\ppcen^*\phi_\ppcen
    \end{array}
                \right), \quad
  M^Q_p \equiv \left(
    \begin{array}{ccc}
      M_{00}^2 & 2 M_{00} M_{01} & M_{01}^2 \\
      M_{00} M_{10} & M_{00} M_{11} + M_{01} M_{10} & M_{01} M_{11} \\
      M_{10}^2 & 2 M_{10} M_{11} & M_{11}^2
    \end{array}
  \right),
\end{displaymath}
where we have abbreviated $(M_p)_{ij}\equiv M_{ij}$; after a straightforward
calculation we obtain
\begin{displaymath}
  (\identity-M^Q_p)^{-1}_{0,0}
    = {1\over 4(\bppl + \bpmi + 2)} + {2\bppl - \bpmi - 2 \over 8\bppl} -
      {\bppl^2 (\bppl + \bpmi + 2) \over
	4(\bppl+\bpmi)\bigl[ (\bppl+1)^2 + (\bpmi+1)^2 \bigr]}.
\end{displaymath}
The integrated autocorrelation $\A_\Msq=(\identity-M^Q_0)^{-1}_{0,0}-1$ is then
given by
\begin{displaymath}
  \A_\Msq \;=\;     {1\over (2 - \orp) \orp m^2} \;
                {\left(
              \begin{array}{c}
                (m^6 + 4 d m^4 + 9 d^2 m^2 + 4 d^3) \orp^4 \\
          -     2 (3 m^6 + 13 d m^4 + 23 d^2 m^2 + 12 d^3) \orp^3 \\
          +     (13 m^6 + 62 d m^4 + 102 d^2 m^2 + 56 d^3) \orp^2 \\
          -     4 (3 m^6 + 16 d m^4 + 28 d^2 m^2 + 16 d^3) \orp \\
          +     4 (m^6 + 6 d m^4 + 12 d^2 m^2 + 8 d^3)
              \end{array}
                \right)
                \over
                \left(
              \begin{array}{c}
	        (m^4 + 4 d m^2 + 8 d^2) \orp^2 \\
	  -     4 (m^4 + 4 d m^2 + 4 d^2) \orp \\
	  +     4 (m^4 + 4 d m^2 + 4 d^2)
              \end{array}
                \right)}.
\end{displaymath}
This equation indicates that without tuning of the overrelaxation parameter we
have $z_\Msq=2$ as expected. To find an optimal tuning we need to minimize
$\A_\Msq$ with respect to $\orp$. Since we are only interested in the asymptotic
behaviour as $m\rightarrow 0$ ($\xi\rightarrow\infty$), it suffices to consider
just the leading terms as this limit is approached; in particular, we have
\begin{displaymath}
  \A_\Msq = {d(2-\orp) \over 2\orp m^2} + {7\orp^2-16\orp+8
    \over 8\orp(2-\orp)} + O(m^4),
\end{displaymath}
which is minimized by
\begin{displaymath}
  \orp^{\min} =  2 - {m \over \sqrt{d}} + O(m^2).
\end{displaymath}
With this tuning the integrated autocorrelation becomes
\begin{displaymath}
  \A_\Msq^{\min} = {\sqrt{d} \over2m} -\half + O(m),
\end{displaymath}
giving $z_\Msq=1$. It is not possible to achieve $z=0$ as in the case of
magnetization.

For completeness we give the expressions for the autocorrelation function.
Following our general strategy we find
\begin{displaymath}
  (\identity-xM^Q_0)^{-1}_{1,1} =
    \sum_{j=1}^3 {a_j\over 1-b_j x} =
    {Q_0 + Q_1 x \over R_0 + R_1 x + R_2 x^2} + {a_3\over 1-b_3 x},
\end{displaymath}
with
\begin{displaymath}
  Q_0 = (\bpl +2)^2 + \bmi(\bmi +4), \qquad
  Q_1 = \left(
    \begin{array}{c}
      - \bmi^2 \bpl^4 \\
      + 4 \bmi^3 \bpl^3 \\
      - (1 + 2\bmi^2(1 + 3\bmi)(1 + \bmi)) \bpl^2 \\
      + 4 (\bmi + 1)^3 (\bmi^2 +\bmi - 1) \bpl \\
      - (\bmi + 1)^4 (\bmi + 2)^2
    \end{array}
    \right),
\end{displaymath}
and
\begin{eqnarray*}
  R_0 &=& (\bpl -\bmi)^2  + 4(1+\bpl+\bmi), \\
  R_1 &=& - R_0 \left(
    \begin{array}{c}
      \bpl^4 \\
      + 4 (1 - \bmi) \bpl^3 \\
      + 2 (3 - 2\bmi + 3\bmi^2) \bpl^2 \\
      + 4 (1 - \bmi - \bmi^2 - \bmi^3) \bpl \\
      + (2 + 4\bmi + 6\bmi^2 + 4\bmi^3 + \bmi^4)
    \end{array}
    \right), \\
  R_2 &=& R_0 (\bpl + \bmi + 1)^4.
\end{eqnarray*}
The coefficients $a_1,b_1,a_2,b_2$ are obtained by inserting these expressions
into equation (\ref{eq:eo-mag-acfcoeff}), while $a_3,b_3$ are given by
\begin{displaymath}
  a_3 = -{2\bpl\bmi\over R_0} \qquad b_3 = (1 + \bpl + \bmi)^2.
\end{displaymath}

\begin{figure}[htb]
  \centerline{\epsfxsize=0.95\textwidth\epsfbox{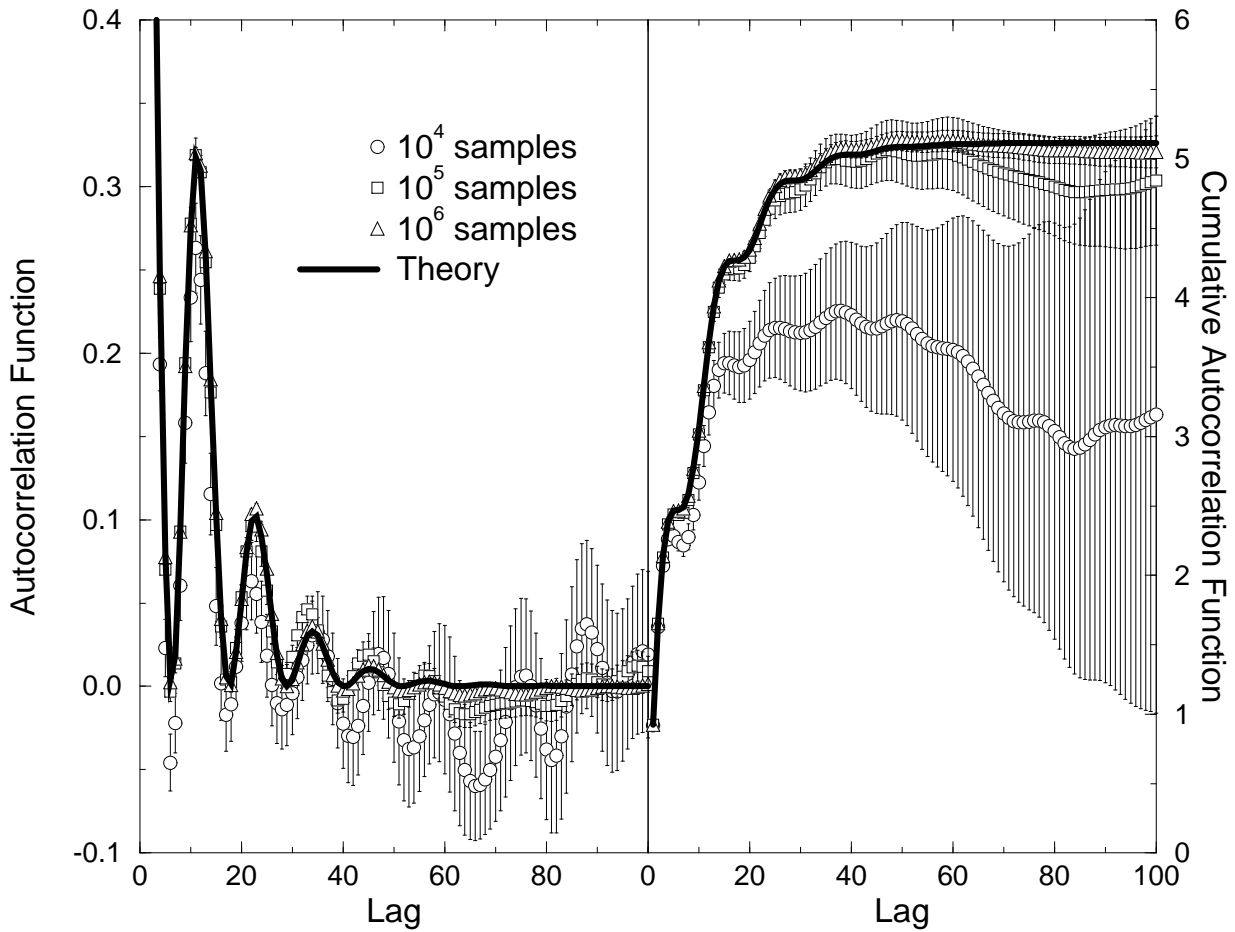}}
  \caption[fig:ms4-2d-eo]{Comparison of theoretical prediction with measured
     values of $\A_\Msq$ for two-dimensional free field theory using even/odd
     updates. The mass $m=0.2$ and the overrelaxation parameter $\orp=0.95$.}
   \label{fig:ms4-2d-eo}
\end{figure}
In Figure~\ref{fig:ms4-2d-eo} we plot the autocorrelation function and the
cumulative autocorrelation function for $\M^2$ in two dimensional free field
theory, together with numerical data for various sample sizes. The error bars on
the numerical data were obtained by binning $n$ measurements into $\sqrt n$
bins. The difficulty in estimating the integrated autocorrelation from small
data sets is immediately apparent.

\subsection{Lexicographic Updates}

In this section we analyze updates in which the variable $\phi_{x-\emu\mu}$ is
always updated before the variable $\phi_x$. Strictly speaking such an updating
scheme does not exist on a finite lattice with periodic boundary conditions; for
example, if we start with variable $\phi_1$ on a one-dimensional lattice with
$L$ sites, then the variable $\phi_{1-\emu1}\equiv\phi_L$ will certainly be
updated after it. Allowing for these violations on the boundaries results in the
class of updates that we call {\em lexicographic\/} after its most common
implementation.

To find the update matrix in this case it is useful to write the linear
stochastic update (\ref{eq:d40}) in a different form. The local update
(\ref{eq:d20}) involves the updated variable and its nearest neighbours; some of
the neighbours, however, might have already been updated in the current sweep.
We may separate the dependence on the ``old'' and ``new'' variables and write
\begin{displaymath}
  \phi' = O\phi + N\phi' + \bar P\pi;
\end{displaymath}
note that the matrix $\bar P = \identity {1\over\freqeqm}\sqrt{\orp(2-\orp)}$ is
different from the matrix $P$ of equation~(\ref{eq:d40}), in fact we have
\begin{equation}
  M = (\identity-N)^{-1}O, \qquad P=(\identity-N)^{-1}\bar P.
  \label{eq:x1}
\end{equation}

It is straightforward to find an explicit form of the matrices $O$ and $N$ for
the lexicographic ordering. We have
\begin{eqnarray*}
  O_{x,y} &=& (1-\orp)\delta_{x,y} +
    {\orp\over\freqeqm^2} \sum_{\mu=1}^d \delta_{x,y-\emu\mu} +
    {1\over L} R_{x,y}, \nonumber \\
  N_{x,y} &=& {\orp\over\freqeqm^2} \sum_{\mu=1}^d \delta_{x,y+\emu\mu} -
    {1\over L} R_{x,y};
\end{eqnarray*}
where we have separated the translationally noninvariant part
\begin{displaymath}
  R_{x,y} \equiv {L\orp\over\freqeqm^2} \sum_{\mu=1}^d
  (\delta_{x,y+\emu\mu}\delta_{x_\mu,1}-\delta_{x,y-\emu\mu}\delta_{x_\mu,L}).
\end{displaymath}
The matrix $R$ represents the violation of our ordering rule on the lattice
boundaries; as such it constitutes only a correction to the translationally
invariant part.

To see this more explicitly and to take advantage of the translational
invariance in the infinite volume limit, it is once again convenient to work in
momentum space. We have
\begin{equation}
  O = O^D + {1\over L}R, \qquad N = N^D - {1\over L} R,
  \label{eq:x4}
\end{equation}
where the translationally invariant parts are given by
\begin{equation}
  O^D_{p,q} = \delta_{p,q} \left(1 - \orp + {\orp\over\freqeqm^2}
    \sum_{\mu=1}^d e^{2\pi iq_\mu/L}\right), \qquad
  N^D_{p,q} = \delta_{p,q} {\orp\over\freqeqm^2}
    \sum_{\mu=1}^d e^{-2\pi iq_\mu/L},
  \label{eq:x5}
\end{equation}
and
\begin{equation}
  R_{p,q} = {\orp\over\freqeqm^2} \sum_{\mu=1}^d \delta_{p,q}^\mu
    \left(e^{-2\pi ip_\mu/L} - e^{2\pi iq_\mu/L}\right);
  \label{eq:x6}
\end{equation}
where in the last equation we have defined $\delta_{p,q}^\mu \equiv
\prod_{\nu\neq\mu} \delta_{p_\nu,q_\nu}$. In momentum space the matrix elements
of the translationally noninvariant contribution are explicitly suppressed by a
factor of $1/L$: this is expected since the violations have support only on the
boundary whose size relative to the bulk volume is given by this factor.

A similar suppression should also be explicit in the form of the update matrix:
in Appendices~\ref{sec:appendix-A} and~\ref{sec:appendix-B} we show that
\begin{equation}
  M = \sum_{l=0}^d {1\over L^l} G^{(l)}
    + O\left({1\over L}\left({\orp\over\freqeqm^2}\right)^L\right).
  \label{eq:x7}
\end{equation}
Here $\orp/\freqeqm^2 < 1/d$ for $m>0$, so that the neglected correction is
exponentially small in lattice size. The matrix elements of $G^{(l)}$ have an
$L$-independent bound and are of the form
\begin{displaymath}
  G^{(l)}_{p,q} = \sum_{\{\mu_1,\ldots,\mu_l\}}\delta_{p,q}^{\mu_1,\ldots,\mu_l}
    \hat G^{\mu_1,\ldots,\mu_l}_{p,q},
\end{displaymath}
with the modified $\delta$-function is defined by
\begin{equation}
  \delta_{p,q}^{\mu_1,\ldots,\mu_l} \equiv
  \prod_{\nu\not\in\{\mu_1,\ldots,\mu_l\}} \delta_{p_\nu,q_\nu}
  \label{eq:x9}
\end{equation}
where $\{\mu_1,\ldots,\mu_l\}$ is a subset of the  integers $\{1,\ldots,d\}$.
The leading term of equation (\ref{eq:x7}) is translation invariant and
corresponds to the situation with no violations of the ordering rule, namely
\begin{equation}
  G^{(0)}_{p,q} = M^D_{p,q}
    = \left[(\identity-N^D)^{-1}O^D\right]_{p,q}
    = \delta_{p,q}
      {\displaystyle{1 - \orp
	+ {\orp\over\freqeqm^2}\sum_{\mu=1}^d e^{2\pi ip_\mu/L}}
      \over \displaystyle{1
	- {\orp\over\freqeqm^2}\sum_{\mu=1}^d e^{-2\pi ip_\mu/L}}}.
  \label{eq:x16}
\end{equation}
The $1/L$ corrections arise from ordering violations on sites which have exactly
one
component having the boundary value ($1$ or~$L$). We still have translational
invariance in the remaining directions, which is expressed by the presence of
$\delta^\nu_{p,q}$: explicitly we have
\begin{equation}
 G^{(1)}_{p,q}
  = \sum_{\mu=1}^d \delta^\mu_{p,q}\hat K^{\mu}_{p,q}
    \left(-\orp + {2\orp\over\freqeqm^2}\sum_{\nu=1}^d
      \cos{2\pi\over L}q_\nu\right),
  \label{eq:x21}
\end{equation}
with $\hat K^{\mu}$ defined in equation (\ref{eq:a65}) of
Appendix~\ref{sec:appendix-A}. Similarly the $1/L^l$ corrections originate with
the sites with $l$ components taking boundary values, and the product of
$\delta$-functions ensures the residual translation invariance in $d-l$
directions.

\subsubsection{$\A_\M$ and $\A_{\M^2}$ at Infinite Volume}

The analysis of autocorrelations becomes straightforward in an infinite volume
because of the diagonal nature of the update matrix. For the magnetization we
have
\begin{equation}
  1 + \A_\M = (\identity-M)^{-1}_{0,0}
    = {1\over 1-M^D_{0,0}}
    = {m^2 + 2d\over\orp m^2} - {d\over m^2}.
  \label{eq:x11}
\end{equation}
Note that this result is identical to that of equation (\ref{eq:eoac}) for
even/odd updates, and choosing
\begin{equation}
  \orp=\orp^c\equiv {m^2+2d\over m^2+d}
  \label{eq:x12}
\end{equation}
we have again $\A_\M\equiv 0$, giving $z_\M=0$. Nevertheless there is an
interesting difference: the autocorrelation function takes the form of a single
exponential, namely
\begin{displaymath}
  \acf_\M(t) = (M^D_{0,0})^t =
    \exp\left(t\ln{1- \orp + d\orp/\freqeqm^2\over 1 - d\orp/\freqeqm^2}\right),
\end{displaymath}
giving the exponential autocorrelation time
\begin{displaymath}
  \ace_\M = \left(\ln{1 - d\orp/\freqeqm^2 \over
    1 - \orp + d\orp/\freqeqm^2}\right)^{-1}.
\end{displaymath}
$\ace_\M$ vanishes at $\orp=\orp^c$, consequently each sweep generates a new
configuration in which the value of magnetization is completely decorrelated
from the previous value.

For $\Msq$ we have similarly
\begin{equation}
  1 + \A_\Msq =
      (\identity-M^Q)^{-1}_{00,00} =
      {1\over 1 - (M^D_{0,0})^2} =
      {(m^2 + 2d - d\orp)^2 \over \orp(2-\orp) m^2 (m^2 + 2d)},
  \label{eq:x15}
\end{equation}
and obviously $\ace_\Msq=\ace_\M/2$. The integrated autocorrelation has a
minimum at $\orp_c$ as defined in (\ref{eq:x12}), where $\A_\Msq$ vanishes,
$z_\Msq=0$, and complete decorrelation is achieved.

We have thus obtained the interesting result that with a lexicographic ordering
of sites it is possible to tune the overrelaxation parameter $\orp$ so that
critical slowing down is completely eliminated. This is true for any quantity
$\Omega(\phi_0)$ which depends only on the zero momentum component of the field:
the Monte Carlo evolution of $\phi_0$ is not coupled to the other modes in the
infinite volume limit, and for one sweep we have
\begin{equation}
  \phi_0 \mapsto \phi_0^\prime = M^D_{0,0}\phi_0 + P^D_{0,0}\pi_0,
\end{equation}
where $P^D={1\over\freqeqm}\sqrt{\orp(2-\orp)}(\identity-N^D)^{-1}$ (from
equation (\ref{eq:x1})). Since $M^D_{0,0}=0$ at $\orp=\orp_c$ we get
\begin{equation}
   \Bigl\langle\Omega(\phi_0)\Omega(\phi^\prime)\Bigr\rangle
   = \Bigl\langle\Omega(\phi_0)\Omega(P^D_{0,0}\pi_0)\Bigr\rangle
   = \Bigl\langle\Omega(\phi_0)\Bigr\rangle
     \Bigl\langle\Omega(\phi^\prime)\Bigr\rangle,
\end{equation}
thus the autocorrelation function as well as the integrated autocorrelation
vanish.

This is in contrast to the situation for even/odd updates: for lexicographic
updates the Fourier modes decouple completely (in the infinite volume limit) and
each mode is updated independently. Consequently, the autocorrelations of any
quantity which depends on a single Fourier component can be tuned in an optimal
fashion. For the even/odd case the modes are coupled in pairs and critical
slowing down for, e.g., $\Msq$ can only be reduced to $z_\Msq=1$ at best.

\subsubsection{Other Quantities at Infinite Volume}

Let us now try to understand in a little more detail what happens at
$\orp=\orp_c$. Consider the ``staggered magnetization'' squared, $\phi_{p_c}^2$,
which is a function of only the highest frequency mode, at $\orp=\orp_c$ we have
\begin{equation}
  1 + \A_{\phi_{p_c}^2} = {(m^2 + 2d)^2 \over m^2(m^2 + 4d)},
\end{equation}
and thus $z_{\phi_{p_c}^2}=2$. In other words at $\orp=\orp_c$ overrelaxation
does not help if we want to measure $\phi_{p_c}^2$: this is not surprising as we
have not sped up the high frequency modes; we took advantage of the fact that
lexicographic updates allowed us to ignore these modes in the situation where we
were not interested in them. On the other hand, if we want to measure
$\phi_{p_c}^2$ we can tune $\orp$ to $\orp_c^\prime = (m^2+2d)/(m^2+3d)$ and
eliminate critical slowing down completely for this quantity. The even/odd
updating scheme does not have this flexibility.

If we are interested in measuring both  $\phi_0^2$ and $\phi_{p_c}^2$ near
criticality we are better off performing two lexicographic simulations without
critical slowing down than one even/odd simulation. This is an extreme case of
what we believe is a generic feature: if we want to optimize simulations with
overrelaxation updates for some set of operators we can usually find a better
solution with lexicographic ordering of sites than with even/odd ordering at
large volumes. The reason is that the even/odd ordering introduces more
constraints into the optimization than lexicographic update by coupling the
modes in pairs.

It is not easy to find an interesting operator that cannot be optimized well
with lexicographic overrelaxation. For the sake of simplicity let us stay in the
domain of quadratic operators of the type given in equation~(\ref{eq:d81}). As
we have already observed, if $K(q)$ is peaked at any single momentum we expect
$z_\Omega\approx 0$. If the operator $\Omega$ couples to both low and high
frequency modes, the contribution to autocorrelations of high frequency modes
relative to low frequency ones will be suppressed by the $\freq_q$ factors in
equation~(\ref{eq:d100}). To illustrate this, consider the simple quantity
$\Omega = \phi_0^2 + \phi_{p_c}^2$; using formula (\ref{eq:d100}) we obtain
\begin{equation}
  1+\A_\Omega = {\displaystyle{1\over 1-(M^D_{0,0})^2} +
    {1\over 1-(M^D_{p_c,p_c})^2}{m^4\over(m^2 + 4d)^2}
      \over\displaystyle 1 + {m^4\over(m^2 + 4d)^2}}.
\end{equation}
We obviously get $z_\Omega=0$ at $\orp=\orp_c$ because the the $1/m^2$ behaviour
of $1/[1-(M^D_{p_c,p_c})^2]$ there is cancelled by the $m^4$ factor from its
coefficient. Examples of important quantities of this type are two point
functions.

According to the above argument, the most ``dangerous'' quantities are the ones
for which $K(q)$ favours high frequency modes but does not completely decouple
from low frequency modes. The obvious such operator is the energy, where the
spectral weights exactly cancel (see equation~(\ref{eq:d10})). In one dimension
we get $z_E=1$ without tuning, $z_E=2$ at $\orp=\orp_c$, and $\orp$ can be tuned
to get $z_E=1/2$ for $\orp = 2 - 2\sqrt{2m}$; optimization for even/odd updates
also leads to $z_E=1/2$.

Note that according to the above discussion, we should get $\zint\approx 0$
for most of the quantities of interest for $\orp=\orp_c$, which optimizes
functions of the zero momentum Fourier component. Furthermore, if we insist on
measuring many operators with different spectra in one simulation, we can
interleave lexicographic sweeps with $\orp=\orp_c$ with sweeps at $\orp$
corresponding to $\zexp=1$. This will give $\zint=0$ for most of the important
operators while ensuring that $\zint\leq1$ for all operators.

\subsubsection{Autocorrelations in a Finite Volume}

While the value of $\zint$ is defined by the infinite volume behaviour, it is
useful to understand the finite size corrections to autocorrelations.
Interestingly, for the magnetization there are no corrections to the infinite
volume result: indeed, we have
\begin{equation}
  (\identity-M)^{-1} = \Bigl[\identity -(\identity-N)^{-1}O\Bigr]^{-1}
    = (\identity-N-O)^{-1}(\identity-N).
  \label{eq:y1}
\end{equation}
From equation (\ref{eq:x4}) we have $\identity -N-O = \identity - N^D - O^D$, so
\begin{displaymath}
  (\identity-M)^{-1}
    = (\identity-M^D)^{-1} + {1\over L}(\identity - N^D - O^D)^{-1}R;
\end{displaymath}
since $R_{0,0}=0$, it follows that
\begin{displaymath}
  1 + \A_\M = (\identity-M)^{-1}_{0,0} = (\identity-M^D)^{-1}_{0,0},
\end{displaymath}
and the result (\ref{eq:x11}) is exact for any volume.

\begin{figure}[htb]
  \centerline{\epsfxsize=0.9\textwidth\epsfbox{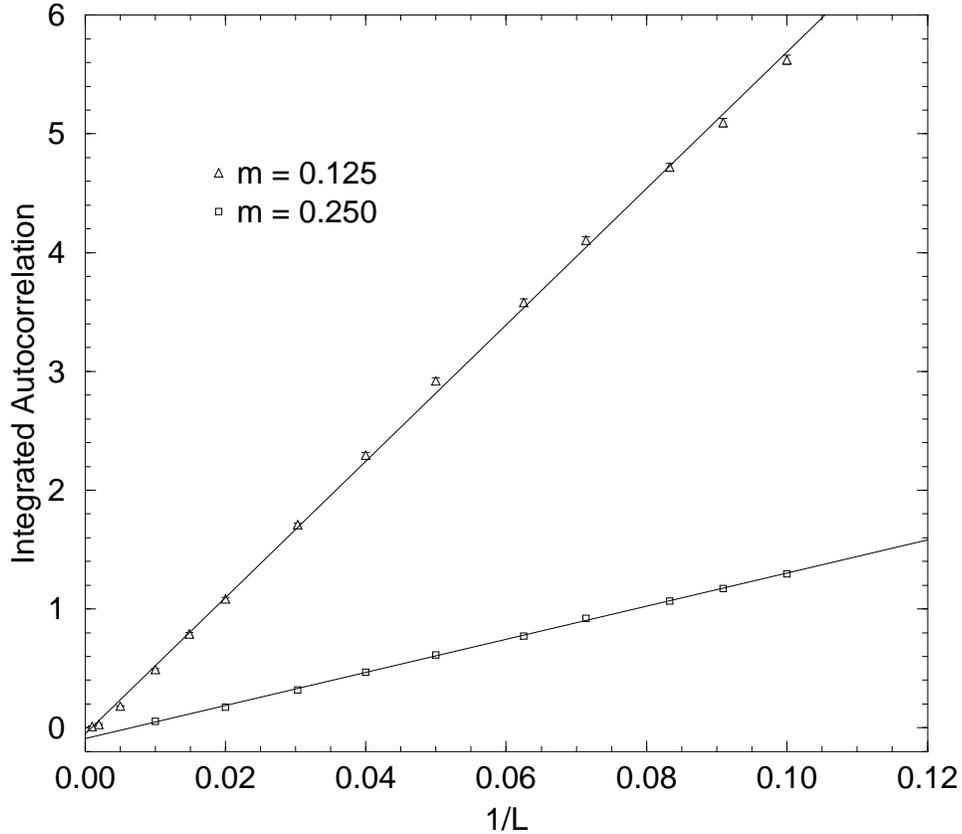}}
  \caption[fig:iac-lex]{$\A_\Msq$ for lexicographic updates in one dimension
    with $m=0.125$ and $0.250$ at $\orp=\orp_c$. The error bars are smaller than
    the symbols.}
  \label{fig:iac-lex}
\end{figure}

In case of $\M^2$ the situation is more complicated. $(\identity - M^Q)^{-1}$ is
expected to have a $1/L$ correction to the translationally invariant part. We
have not been able to evaluate the correction in the closed form and we have no
reason to believe that the zero momentum matrix element vanishes; in fact, our
numerical experiments in one dimension confirm that the correction is of order
$1/L$ as is illustrated in Figure~\ref{fig:iac-lex}. The data shows a very small
curvature at large values of $L$, which corresponds to the small negative
intercepts for the linear fits; we do not understand the cause of this subtle
effect.

\subsubsection{Random Lexicographic Updates}

On physical grounds it is obvious that the effect of surface terms on behaviour
of $\A_{\Omega}$ for any quantity $\Omega$ will go to zero, typically as $1/L$,
in the limit of large volumes. A periodic lattice does not have any boundaries,
it is only the updating scheme that introduces a boundary as the collection of
points at which the lexicographic property is violated; for this reason
expressions describing autocorrelations do not have exact translational
invariance. An obvious way to enhance translational invariance is to randomize
the choice of starting point for each sweep through the lattice. For this case
we are able to provide a relatively simple direct proof that surface effects
only give small corrections to the integrated autocorrelation even for quadratic
operators.

The boundary is completely specified by the site that is updated first in the
sweep; if $x \equiv (x_1,x_2,\cdots,x_d)$ is this first site then a site $y$
belongs to the boundary if $y_j = x_j$ or $y_j = x_{j+L-1}$ for at least one
$j$. We thus need to give every lattice site an equal chance of being the
starting point of the lexicographic chain.

Consider the representation of lexicographic updates in terms of the local
update matrices~(\ref{eq:d40})
\begin{displaymath}
  M = M_{x_{L^d}} M_{x_{L^d-1}} \cdots M_{x_2} M_{x_1},
\end{displaymath}
where the sequence $x_1,x_2,\ldots,x_{L^d}$ defines the lexicographic ordering.
If we relabel the sites by a lattice translation $x\mapsto x-a$ and apply the
same lexicographic update in the new coordinates then in the old coordinates
this corresponds to choosing the initial site to be $x_1+a$ and the update
matrix to be
\begin{displaymath}
  M^{(a)} = M_{x_{L^d}+a} M_{x_{L^d-1}+a} \cdots M_{x_2+a} M_{x_1+a}.
\end{displaymath}
The matrix $M_{x+a}$ can be written in the form
\begin{displaymath}
  M_{x+a} = S^{(a)} M_x S^{(-a)}, \qquad S^{(a)}_{x,y} \equiv \delta_{x,y+a}.
\end{displaymath}
Using the composition property $S^{(a)}S^{(b)}=S^{(a+b)}$ we have
\begin{displaymath}
  M^{(a)} = S^{(a)} M S^{(-a)}.
\end{displaymath}

We now propose the updating scheme where before any given sweep, this
relabelling of sites is performed at random. In other words, the translation
vector $a$ will be a random variable, uniformly distributed over $\Z_L^d$. The
linear update matrix averaged over these translations is given by
\begin{displaymath}
  \bar M = {1\over L^d} \sum_{a\in \Z_L^d} S^{(a)} M S^{(-a)}.
\end{displaymath}
In Fourier space we have
\begin{displaymath}
  S^{(a)}_{p,q} = \delta_{p,q} e^{-i{2\pi\over L}p\cdot a},
    \qquad\mbox{and}\qquad\bar M_{p,q} = \delta_{p,q} M_{p,q};
\end{displaymath}
as expected, the linear update matrix retains only the diagonal part of $M$.

The quadratic update matrix in the $\{\phi_p^*\phi_q\}$ basis is similarly given
by
\begin{displaymath}
  \bar M^Q = {1\over L^d} \sum_{a\in \Z_L^d}
    (S^{(a)} M S^{(-a)})^* \otimes (S^{(a)} M S^{(-a)}).
\end{displaymath}
Performing the sum over translations yields
\begin{displaymath}
  \bar M^Q_{pq,rs} = M_{p,r}^*M_{q,s} \delta_{p-q,r-s},
\end{displaymath}
implying that the quadratic monomial $\phi_p^*\phi_q$ couples only to monomials
$\phi_r^*\phi_s$ satisfying $p-q=r-s$. Restricting ourselves to the sector with
$p-q=0$ and defining $\Phi_p \equiv \phi_p^*\phi_p$ we have
\begin{displaymath}
  \Phi' = \bar M^{Q,0} \Phi, \qquad\mbox{with}\qquad
  \bar M^{Q,0}_{p,q} = M_{p,q}^*M_{p,q};
\end{displaymath}
here $\bar M^{Q,0}$ is the corresponding $L^d\times L^d$ block of the quadratic
update matrix given by the Hadamard product $M^*\cdot M$. We are interested in
showing that
\begin{equation}
  1+\A_{\M^2} = (\identity - M^*\cdot M)^{-1}_{0,0} \to
    (\identity - {M^D}^*\cdot M^D)^{-1}_{0,0} \qquad\mbox{as $L\to\infty$.}
  \label{eq:y16}
\end{equation}

Let us start by writing the matrix $M$ explicitly as~(\ref{eq:x7})
\begin{displaymath}
  M = M^D + \sum_{l=1}^d {1\over L^l} G^{(l)},
\end{displaymath}
where we have absorbed the exponentially small corrections into the matrices
$G^{(l)}$; we then have a decomposition
\begin{displaymath}
  \identity - M^*\cdot M = \identity - {M^D}^*\cdot M^D - E^{(1)} - E^{(2)}
    \equiv \identity - {M^D}^*\cdot M^D - E,
\end{displaymath}
where
\begin{eqnarray}
  E^{(1)}_{p,q} &=& M^D_{p,q} \sum_{l=1}^d {1\over L^l} {G^{(l)}_{p,q}}^*
    + {M^{D}_{p,q}}^* \sum_{l=1}^d {1\over L^l} G^{(l)}_{p,q} \\
  E^{(2)}_{p,q} &=&
      \sum_{j,l=1}^d {1\over L^{l+j}} {G^{(l)}_{p,q}}^* G^{(j)}_{p,q}
    = \sum_{j,l=1}^d\sum_{\scriptstyle \{\mu_1,\ldots,\mu_l\} \atop
      \scriptstyle \{\nu_1,\ldots,\nu_j\}}
      {1\over L^{l+j}} \delta_{p,q}^{\mu_1,\ldots,\mu_l}
	\delta_{p,q}^{\nu_1,\ldots,\nu_j}
	\left.{\hat G}^{\mu_1,\ldots,\mu_l}_{p,q}\right.^*
	{\hat G}^{\nu_1,\ldots,\nu_j}_{p,q}.
  \label{eq:y28}
\end{eqnarray}
We now argue that the ``error'' matrix $E$ is small at large volumes in the
sense that its matrix norm tends to zero in that limit. To show this we use the
maximum sum row norm $\|A\| \equiv \max_p \sum_q |A_{p,q}|$. Consider first
$E^{(1)}$, which is a diagonal matrix since $M^D$ is diagonal; the matrix
elements of $M^D$ and $G^{(l)}$ are bounded by $L$-independent constants, thus
there is a constant $C_1$ such that $\|E^{(1)}\|\le C_1/L$ for all $L$. Turning
to $E^{(2)}$, since there are an $L$-independent number of terms in
equation~(\ref{eq:y28}), it suffices to consider the norm of a general term in
that sum. We have
\begin{equation}
  \sum_q {1\over L^{l+j}} \delta_{p,q}^{\mu_1,\ldots,\mu_l}
    \delta_{p,q}^{\nu_1,\ldots,\nu_j}
      \left| \left.\hat G^{\mu_1,\ldots,\mu_l}_{p,q}\right.^*
	  \hat G^{\nu_1,\ldots,\nu_j}_{p,q}\right|
	< {C^2 \over L^{l+j-k}} \le {C^2 \over L},
  \label{eq:y30}
\end{equation}
where $k$ is the cardinality of the set $\{\mu_1,\ldots,\mu_l\} \intersection
\{\nu_1,\ldots,\nu_j\}$; we have again used the fact that the matrix elements of
the matrices $\hat G^{\mu_1,\ldots,\mu_l}$ have an $L$-independent bound $|\hat
G^{\mu_1,\ldots,\mu_l}_{p,q}|< C$. The last inequality in (\ref{eq:y30}) follows
from the fact that $k\le\min(l,j)$ and consequently $l+j-k \geq 1$. We have thus
established that $\|E^{(2)}\|\le C_2/L$ for all $L$. Putting the two bounds
together we have
\begin{displaymath}
  \|E\| \le {C_1+C_2 \over L} \equiv  {C_3 \over L}.
\end{displaymath}

As the final ingredient we use the standard formula for the error in the
inverse, namely
\begin{equation}
  \left\|(A-E)^{-1}-A^{-1}\right\|\leq{\|A^{-1}\|^2\over1-\|A^{-1}E\|}\|E\|;
  \label{eq:y32}
\end{equation}
this is valid\footnote{The derivation involves the expansion of $(A-E)^{-1}$ in
a Neumann series.} for any matrix $A$ and ``error'' $E$ and any norm in which
$\|A^{-1}E\|<1$.

Let $A\equiv\identity - {M^D}^*\cdot M^D$, then $\|A^{-1}\|$ has an
$L$-independent upper bound and $\lim_{L\to\infty}\|E\|=0$, so we have
$\|A^{-1}E\| \leq \|A^{-1}\|\cdot\|E\| < 1$ for sufficiently large $L$. Formula
(\ref{eq:y32}) is thus applicable, and
\begin{equation}
  \left\|(\identity-M^*\cdot M)^{-1}-(\identity-{M^D}^*\cdot M^D)^{-1}\right\|
    \leq{C_4 \over L},
  \label{eq:y44}
\end{equation}
for sufficiently large $L$; the required result (\ref{eq:y16}) follows.

\begin{figure}[htb]
  \centerline{\epsfxsize=0.9\textwidth\epsfbox{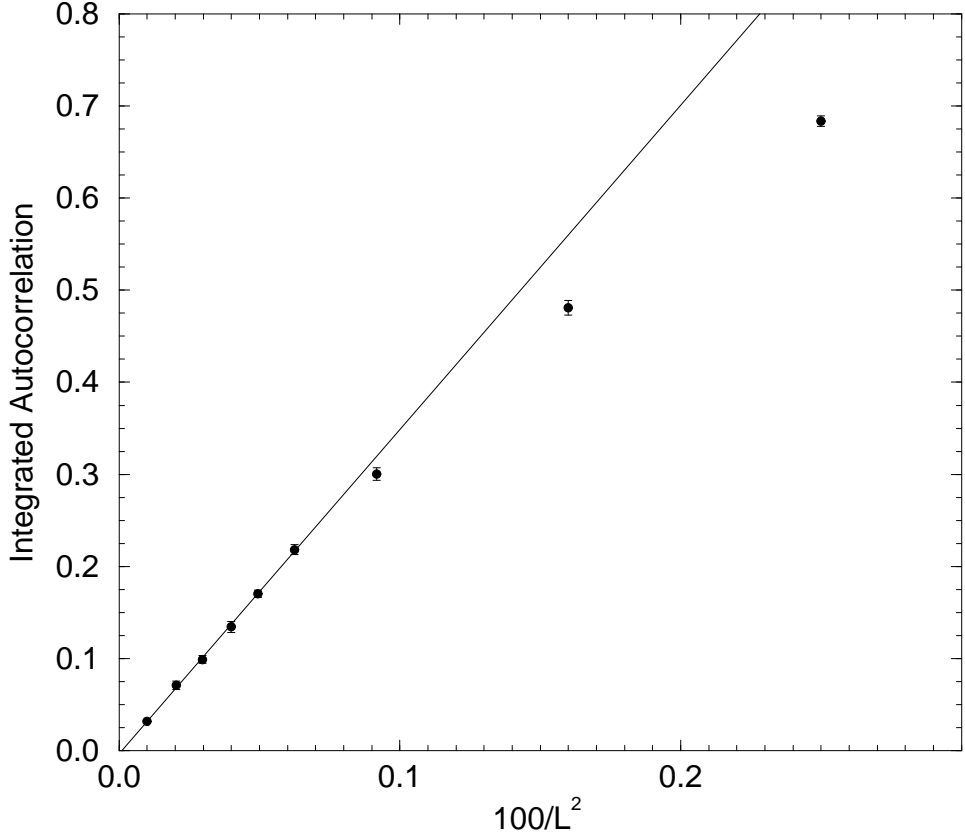}}
  \caption[fig:iac-ranlex]{$\A_\Msq$ for random lexicographic updates in one
    dimension with $m=0.25$ at $\orp=\orp_c$. The $1/L^2$ fit was made to the
    leftmost six points, but not constrained to go through the origin.}
  \label{fig:iac-ranlex}
\end{figure}

In the infinite volume limit $\A_\M^2$ is the same as for the standard
lexicographic update (\ref{eq:x15}). Whilst equation (\ref{eq:y44}) tells us
that the leading finite volume correction is at worst $O(1/L)$, it is actually
$O(1/L^2)$ at $\orp=\orp_c$. Indeed, the matrix elements of $E^{(2)}$ are
supressed by $1/L^2$ and consequently the $O(1/L)$ part is determined solely by
$E^{(1)}$, which is diagonal. Using the explicit form (\ref{eq:x21}) of matrix
$G^{(1)}$ this leading correction can be evaluated, and the correction vanishes
at $\orp=\orp_c$. The $1/L^2$ behaviour of the correction is in accord with
numerical results, as shown in Figure~\ref{fig:iac-ranlex}.

\subsection{Random Updates}

Finally consider an updating scheme in which the sites to be updated are chosen
at random. A sequence of $L^d$ such local updates will be called a ``sweep''.

For a single site update we use the update matrix $M_z$ given in
equation~(\ref{eq:d35}) with the updated site $z$ being a random variable,
uniformly distributed over $\Z_L^d$. In Fourier space we obtain
\begin{displaymath}
  (M_z)_{p,q} = \delta_{p,q} + {1\over L^d} e^{-2\pi i(p-q)z/L} f(q),
\end{displaymath}
with
\begin{displaymath}
  f(q) =
  \orp\biggl[{2\over\freqeqm^2}\sum_{\mu=1}^d \cos{2\pi\over L}q_\mu - 1\biggr].
\end{displaymath}
For autocorrelations of quadratic operators we define the quadratic elementary
update matrix $M_z^Q$ in basis $(\phi_p^*\phi_q)$, namely
\begin{displaymath}
  (M_z^Q)_{pq,rs} = (M_z^*\otimes M_z)_{pq,rs} = (M_z)_{p,r}^* (M_z)_{q,s}.
\end{displaymath}

\subsubsection{Autocorrelations for $\M$}

The linear update matrix for a random update sweep, $M$, averaged over the
independent random variables $z_i$ is given by
\begin{displaymath}
  M = \prod_{i=1}^{L^d}\biggl({1\over L^d}\sum_{z_i=1}^{L^d}M_{z_i}\biggr)
    = \biggl({1\over L^d} \sum_{z=1}^{L^d} M_z\biggr)^{L^d}.
\end{displaymath}
Explicitly, we have
\begin{displaymath}
  M_{p,q} = \delta_{p,q} \left(1 + {f(q)\over L^d}\right)^{L^d}
    = \delta_{p,q} e^{f(q)}
      \left(1 - {f(q)^2\over 2L^d} + O\left({1\over L^{2d}}\right)\right),
\end{displaymath}
giving the integrated autocorrelation
\begin{displaymath}
  1 + \A_\M = [\identity - M]^{-1}_{0,0}
    = {1\over1 - \exp\left(-{\orp m^2\over m^2 + 2d}\right)}
      + {\orp^2 m^4 \over L^d 8 (m^2 + 2d)^2
	\sinh^2\left({\orp m^2\over 2(m^2 + 2d)}\right)}
      + O\left({1\over L^{2d}}\right).
\end{displaymath}
In the small mass limit we have
\begin{displaymath}
  1 + \A_\M = {2d\over\orp m^2} + O(m^0),
\end{displaymath}
indicating that $z_\M=2$ regardless of the choice of $\orp$. Even though the
modes have completely decoupled and the autocorrelation function is a single
exponential, the random choice of the site to be updated completely destroys the
coherence and the algorithm cannot be tuned to reduce the critical slowing down.

\subsubsection{Autocorrelations for $\Msq$}

The quadratic update matrix $M^Q$ averaged over the random variables $z_i$ is
\begin{equation}
  M^Q = \prod_{i=1}^{L^d} \biggl({1\over L^d}\sum_{z_i=1}^{L^d}M^Q_{z_i}\biggr)
    = \biggl({1\over L^d} \sum_{z=1}^{L^d} M^Q_z\biggr)^{L^d}.
  \label{eq:r9}
\end{equation}
Straightforward evaluation gives
\begin{displaymath}
  {1\over L^d} \sum_{z=1}^{L^d} M^Q_z
    = \identity + {1\over L^d} \left(D^Q + {1\over L^d}C^Q\right),
\end{displaymath}
with
\begin{displaymath}
  D^Q_{pq,rs} = \delta_{p,r}\delta_{q,s}\bigl(f(r) + f(s)\bigr), \qquad
  C^Q_{pq,rs} = \delta_{p-r,q-s}f(r)f(s).
\end{displaymath}
Inserting this into equation~(\ref{eq:r9}) yields
\begin{displaymath}
  M^Q = \exp(D^Q) + O\left({1\over L^d}\right);
\end{displaymath}
we have not evaluated the $O(1/L^d)$ correction in this case. In the infinite
volume limit we obtain
\begin{displaymath}
  [\identity - M^Q]^{-1}_{pq,rs}
    = {\delta_{p,r}\delta_{q,s}\over 1 - e^{f(r)+f(s)}},
\end{displaymath}
and the integrated autocorrelation of $\Msq$ is given by
\begin{displaymath}
  1 + \A_\Msq
    = {1\over 1 - \exp\left(-{2\orp m^2 \over m^2 + 2d}\right)}
    \approx {d \over \orp m^2} + O(m^0) \qquad\mbox{for $m\to0$}.
\end{displaymath}
We therefore find that again that the algorithm with random updates cannot be
tuned to reduce critical slowing down below $z_{\M^2}=2$.

\section{Conclusions} \label{sec:conclusions}

The principal lesson which we may draw from this analysis is that it is possible
to reduce the cost of a Local Hybrid Monte Carlo computation by a judicious
choice of the order in which the variables are updated, as well as a careful
tuning of the amount of randomness which is introduced into the system. While
this differs from some previous claims, it is similar to the situation for
deterministic Gauss-Seidel linear equation solvers.

The relevant autocorrelation for the determination of the cost is the {\em
integrated autocorrelation}, and it is in terms of this autocorrelation that we
define the critical exponent $\zint$. Furthermore, we define $\zint$ by
$\A\propto\xi^\zint$ as $L,\xi\to\infty$ and $\xi/L\to0$. This is to be
contrasted with a finite size scaling analysis in which the limits are taken in
a different order. For a zero temperature quantum field theory our choice of
limits seems more natural.

Most local algorithms are special cases of LHMC, so our analysis is of fairly
general applicability if we assume that the behaviour of autocorrelations in
asymptotically free interacting theories is similar to that of free field
theory. We intend to study this empirically in a future publication.

Of course local algorithms are directly applicable only to theories with local
bosonic actions, but they are used within some dynamical fermion algorithms such
as that proposed by L\"uscher~\cite{luescher94a}.

One issue of practical importance which has not been addressed at all in this
paper is the difficulty of implementing the lexicographic scheme on parallel
computers. Nevertheless it may well be advantageous to use a local lexicographic
scheme on parallel computers, as has been suggested for the case of
preconditioning conjugate gradient linear equation solvers~\cite{frommer96a}.

\section*{Acknowledgements}
We would like to thank Stephen Adler, Andreas Frommer, Robert Mendris, Mike
Navon, and Yousef Saad for useful discussions. This research was supported by by
the U.S. Department of Energy through Contract Nos.~DE-FG05-92ER40742
and~DE-FC05-85ER250000.

\appendix\section{Calculation of $M$ in the Lexicographic Scheme}
\label{sec:appendix-A}

Our aim in this section is to show that the update matrix $M$ for lexicographic
updates has the structure as given in equation (\ref{eq:x7}), with the leading
term of equation~(\ref{eq:x16}) and subleading term of equation~(\ref{eq:x21}).
We shall prove this by induction in the number of dimensions $d$ for fixed
lattice size~$L$.  The update matrix is a function of the mass and the
overrelaxation parameter as well as $d$ and $L$, $M^{(d)}=M^{(d)}(m,\orp)$. The
induction step assumes that $M^{(d)}$ is known for all $m>0$ and $0<\orp<2$ and
relates  $M^{(d+1)}(m,\orp)$ to~$M^{(d)}(m',\orp)$.

Consider the vector of field variables $\phi\equiv (\phi_1, \phi_2, \ldots,
\phi_L)$ in $d+1$ dimensions, where we only keep the index corresponding to the
new dimension $d+1$, that is, $\phi_x$ is itself a vector of field variables in
$d$ dimensions. We first find the matrix $M^{(d+1)}_z$, corresponding to
lexicographic update of the variables $\phi_z$ only. Separating the dependence
on the ``old'' and ``new'' variables in this subspace as usual, we obtain
\begin{displaymath}
  \phi_z^\prime
    = O^{(d)}\phi_z + N^{(d)}\phi_z^\prime
      + \identity{\orp\over\freqeqm^2}(\phi_{z+1}-\phi_{z-1}),
\end{displaymath}
or
\begin{displaymath}
  \phi_z^\prime
    = M^{(d)}\phi_z
      + (\identity - N^{(d)})^{-1}{\orp\over\freqeqm^2}(\phi_{z+1}-\phi_{z-1}).
\end{displaymath}
We emphasise that the system under consideration is $d+1$ dimensional, and so
$\freqeqm^2=m^2+2(d+1)$; furthermore the update matrix $M^{(d)}$ for a
$d$-dimensional subspace is the same as that for a $d$-dimensional system with
mass ${m'}^2=2+m^2$. For convenience we define a variable $t\equiv
\orp/\freqeqm^2$; and we consider $N^{(d)}$ to be a function of $t$, and
$O^{(d)}$ and $M^{(d)}$ to be functions of both $t$ and~$\orp$. We can thus
write
\begin{displaymath}
  \left[M_z^{(d+1)}\right]_{x,y}
    = \identity\delta_{x,y}
      + \delta_{y,z} \Bigl[(M^{(d)}-\identity)\delta_{x,y}
	+ t(\identity - N^{(d)})^{-1} (\delta_{x+1,y} + \delta_{x-1,y})\Bigr].
\end{displaymath}

The full update matrix is the product (in lexicographic order) of the subspace
update matrices. In particular,
\begin{displaymath}
  M^{(d+1)} = M_L^{(d+1)} M_{L-1}^{(d+1)} \ldots M_1^{(d+1)}
    = (M_L^{(d+1)} S)^L \equiv (\hat M_L^{(d+1)})^L,
\end{displaymath}
where $S_{x,y}\equiv \identity\delta_{x,y-1}$ represents a translation
in the new coordinate $d+1$.

In Fourier space it is convenient to introduce $\alpha\equiv e^{2\pi i/L}$, so
we have
\begin{displaymath}
  \left[{\hat M}_L^{(d+1)}\right]_{p,q}
    = \left\{\identity\delta_{p,q}
      + {1\over L} \Bigl[M^{(d)} - \identity
	+ t(\identity - N^{(d)})^{-1} (\alpha^q + \alpha^{-q})\Bigr]
      \right\} \alpha^q.
\end{displaymath}
Note that we only show the indices corresponding to the new dimension. Writing
$T\equiv t(\identity - N^{(d)})^{-1}$  one can show by induction that the matrix
elements of the $l$th power for $l<L$ are given by
\begin{displaymath}
  \left[\left.{\hat M}_L^{(d+1)}\right.^l\right]_{p,q}
    = \left\{\identity\delta_{p,q}\alpha^{q(l-1)}
      + {1\over L}{\cal P}_{l-1}(\alpha^p,\alpha^q,T)
	\Bigl[M^{(d)} - \identity  + t(\identity - N^{(d)})^{-1}
	  (\alpha^q + \alpha^{-q})\Bigr]\right\} \alpha^q,
\end{displaymath}
with
\begin{displaymath}
  {\cal P}_l(x_1,\ldots,x_n)
    = \sum_{m_1,\ldots,m_n\ge0 \atop m_1+\cdots+m_n=l}
      x_1^{m_1}\ldots x_n^{m_n},
\end{displaymath}
which satisfies the following useful identities
\begin{eqnarray*}
  {\cal P}_l(x_1,\ldots,x_n)
    &=& \sum_{j=0}^l x_n^j {\cal P}_{l-j}(x_1,\ldots,x_{n-1}) \\
  {\cal P}_l(x_1,\ldots,x_n)
    &=& x_n {\cal P}_{l-1}(x_1,\ldots,x_n) + {\cal P}_l(x_1,\ldots,x_{n-1}) \\
  {\cal P}_l(x,y) &=& {x^{l+1} - y^{l+1}\over x-y}.
\end{eqnarray*}
The $L$th power has one extra term giving
\begin{equation}
  \left[M^{(d+1)}\right]_{p,q}
    = \identity\delta_{p,q}
      + {\alpha^q \over L} \Bigl[{\cal P}_{L-1}(\alpha^p,\alpha^q,T) + T\Bigr]
	\Bigl[M^{(d)} - \identity  + t(\identity - N^{(d)})^{-1}
	  (\alpha^q + \alpha^{-q}) \Bigr].
  \label{eq:a20}
\end{equation}
We can obtain a simpler recursion relation for $(\identity - N)^{-1}$, which is
related to $M$ through multiplication by the diagonal matrix; indeed, using
equations (\ref{eq:y1}) and (\ref{eq:x4}) we get from~(\ref{eq:a20})
\begin{equation}
  \left[\Bigl(\identity - N^{(d+1)} \Bigr)^{-1}\right]_{p,q}
    = {\alpha^q \over L} \Bigl[{\cal P}_{L-1}(\alpha^p,\alpha^q,T) + T\Bigr]
     \Bigl(\identity - N^{(d)} \Bigr)^{-1}.
  \label{eq:a25}
\end{equation}
Finally we need to transform ${\cal P}_{L-1}(\alpha^p,\alpha^q,T)$ into an
explicit form; straightforward algebraic manipulation leads to
\begin{equation}
  {\cal P}_{L-1}(\alpha^p,\alpha^q,T) =
    \delta_{p,q}L(\alpha^p\identity - T)^{-1} -
    T(\alpha^p\identity-T)^{-1}(\alpha^q\identity-T)^{-1}(\identity-T^L).
  \label{eq:a30}
\end{equation}
Recalling the definition of $T$ we have also
\begin{equation}
  (\alpha^p\identity - T)^{-1}
    = {\PB\over 1-t\PB} \left(\identity - {1\over 1-t\PB}N^{(d)}\right)^{-1}
      (\identity-N^{(d)}).
  \label{eq:a35}
\end{equation}
It is important to note here that the inverse on the right of this equation
exists: we know that $(\identity - N^{(d)})^{-1}$ exists for all $|t|<1/d$,
where $t$ enters as a multiplicative factor in definition of $N^{(d)}$ (see
equations (\ref{eq:x4})---(\ref{eq:x6})). However, in $d+1$ dimensions $t$
satisfies a more stringent constraint, namely $t < 1/(d+1)$, and consequently
\begin{displaymath}
  \left|{t\over 1-t\PB}\right| \le \left|{t\over 1-t}\right| < {1\over d},
\end{displaymath}
implying the existence of $(\identity - {1\over 1-t\PB}N^{(d)})^{-1}$ for
all~$p$. Inserting relations (\ref{eq:a30}) and (\ref{eq:a35}) into equation
(\ref{eq:a25}) we get an explicit recursion relation
\begin{eqnarray}
  \Bigl[\identity &-& N^{(d+1)} \Bigr]^{-1}_{p,q}
    = \delta_{p,q}{1\over 1-t\PB}
      \left(\identity-{1\over 1-t\PB}N^{(d)} \right)^{-1} \nonumber \\
    &-& {t\over L}\Bigl(\identity - T^L\Bigr){\PB \over (1-t\PB)(1-t\QB)}
      \left(\identity-{1\over 1-t\PB}N^{(d)}\right)^{-1}
      \left(\identity-{1\over 1-t\QB}N^{(d)}\right)^{-1} \nonumber \\
    &+& {t\over L} \QQ \left(\identity-N^{(d)}\right)^{-2}.
  \label{eq:a40}
\end{eqnarray}

In principle this allows us to calculate the update matrix in any dimension.
Consider building a $1$-dimensional update matrix from single site updates
($d=0$); Obviously, $N^{(0)}\equiv 0$ and the recursion relation gives the exact
result
\begin{equation}
  \left[\identity - N^{(1)}\right]^{-1}_{p,q}
    = {\delta_{p,q} \over 1-t\PB} + {t \over L}
      \left[\QQ- {\PB(1-t^L)\over (1-t\PB)(1-t\QB)}\right].
  \label{eq:a45}
\end{equation}
Note that the above result has the expected structure and the leading term
exactly corresponds to $[\identity-{N^{(1)}}^D]^{-1}$ with ${N^{(d)}}^D$
denoting a diagonal part of $N^{(d)}$, defined in~(\ref{eq:x5}). Using $N^{(1)}$
we can calculate  $N^{(2)}$ etc. The only problem from the practical point of
view is the calculation of $T^L=t^L[(\identity-N^{(d)})^{-1}]^L$. However, since
$t<1$ this term is exponentially small in $L$ and can be neglected at large
volumes. Similarly, the exponentially small terms will be generated (and can be
neglected) from matrix multiplications in the recursion relation when using,
e.g., the Poisson resummation formula for the matrix elements of the product.

While the above program can be carried out to arbitrary number of dimensions, it
is sufficient for our purposes to deduce the structure that follows from
equations (\ref{eq:a40}) and (\ref{eq:a45}). Indeed, one can easily verify the
following conclusions:
\begin{enumerate}
  \item In any number of dimensions we have
  \begin{displaymath}
    \left[ \identity - N \right]^{-1} =
     \sum_{l=0}^d {1\over L^l} K^{(l)} + O\left({1\over L}t^L\right).
  \end{displaymath}
  The matrix elements of $K^{(l)}$ are bounded by an $L$-independent bound and
  have the form
  \begin{displaymath}
    K^{(l)}_{p,q} = \sum_{\{\mu_1,\ldots,\mu_l\}}
	      \delta_{p,q}^{\mu_1,\ldots,\mu_l}
	      \hat K^{\mu_1,\ldots,\mu_l}_{p,q},
  \end{displaymath}
  where the modified delta function is defined in equation (\ref{eq:x9}), and
  $\{\mu_1,\ldots,\mu_l\}$ represents any subset of integers $\{1,\ldots,d\}$
  with $l$ elements. This follows from the (\ref{eq:a45}) and the fact that the
  above structure is invariant with respect to recursion relation
  (\ref{eq:a40}). Note that the exponentially small terms could also be included
  in matrices $K^{(l)}$, but we keep them separately for the practical reasons
  mentioned above.

  \item The invariant form of the leading term is given by
  \begin{displaymath}
  \hat K_{p,q} =
    {1 \over \displaystyle{1 - t\sum_\mu\PBM}} ,
  \end{displaymath}
  which corresponds to $[\identity-N^D]^{-1}$ as expected.

  \item The invariant form of the first finite volume correction is
  \begin{equation}
  \hat K^{\mu}_{p,q} = t \left[{\QQM \over
      \displaystyle{\Bigl(1-t\sum_{\nu\neq\mu} \PBN\Bigr)
		    \Bigl(1-t\sum_{\epsilon\neq\mu} \QBE\Bigr)}} -
				      {\PBM \over
      \displaystyle{\Bigl(1-t\sum_\nu \PBN\Bigr)
		    \Bigl(1-t\sum_\epsilon \QBE\Bigr)}} \right].
      \label{eq:a65}
  \end{equation}
  The corresponding forms for the update matrix $M$ are given in equations
  (\ref{eq:x7}-\ref{eq:x21}).
\end{enumerate}

\section{Boundedness of Finite Volume Corrections}
\label{sec:appendix-B}

In this appendix we show that the surface corrections for the lexicographic
updating scheme do indeed go to zero as the inverse of the lattice size. For the
case of quadratic operators we shall assume that the infinite volume limit of
$\det(\identity - M^*\otimes M)\neq0$: we have not yet found a simple proof of
this, although it would follow trivially from a proof of the ergodicity of the
LHMC algorithm in an infinite volume.

The basic idea of the proof is to expand about the infinite volume limit, and
thus our proof is naturally expressed in the language of functional analysis.
Our argument is essentially that the update operator $M$ is a compact
(completely continuous) operator on the Hilbert space $L^2([0,2\pi])$ --- in
fact in one dimension it is a Hilbert-Schmidt kernel --- and that the finite
volume results may be expanded about the infinite volume ones using the Poisson
resummation formula. We have chosen to prove our results following the original
method of Fredholm, as this seems to be the most direct
approach~\cite{riesz55a}.

We start by proving a simple bound on determinants:
\begin{lemma}[Hadamard]
  For any matrix $A:l^2_k(\C)\to l^2_k(\C)$ $$|\det A|\leq\prod_i \|A_i\|$$
  where $A_i$ is a row of $A$ and $\|A_i\|^2 = \sum_j|A_{ij}|^2$ is its norm.
\end{lemma}
\begin{proof}
  If $A$ has any row which is zero then the inequality is trivially satisfied.
  Otherwise construct a new matrix $B$ whose rows are $B_i=A_i/\|A_i\|$, then
  the bound is equivalent to $|\det B|\leq1$ for all complex matrices whose rows
  are normalized. If the rows of $B$ are not all orthogonal then there are two
  rows $u$ and $v$ such that $|(u,v)|>0$, $\|u\|=\|v\|=1$. If we replace the row
  $v$ by $v'\equiv v-(u,v)u$ then the determinant is unchanged, and $(u,v')=0$.
  On the other hand $\|v\|^2= \|v'+(u,v)u\|^2= \|v'\|^2+ |(u,v)|^2\|u\|^2$, so
  $\|v'\|^2= 1-|(u,v)|^2<1$; thus replacing the row $v'$ by $v''\equiv
  v'/\|v'\|$ increases the determinant by a factor of $1/\|v'\|>1$, and all the
  rows of the resulting matrix have unit length. This proves that the maximum
  value of the determinant must occur when the rows are orthonormal. If the rows
  of $B$ are orthonormal then $BB^\dagger=1$ and hence $\det BB^\dagger=\det
  B\det B^\dagger=\det B(\det B)^*=|\det B|^2=1$.
\end{proof}

\begin{corollary}
  If all the elements of the matrix $A$ are bounded, $|A_{ij}| \leq C$, then
  $|\det A| \leq (C\sqrt k)^k$.
\end{corollary}

\begin{theorem}[Fredholm] \label{theorem:fredholm}
  Let $\Delta:l^2_L(\C)\to l^2_L(\C)$ be a matrix all of whose elements are
  bounded by some ($L$-independent) constant $|\Delta_{ij}|\leq C$; then
  $\left|\det\left(\identity + {1\over L}\Delta\right)\right|\leq C'$ where $C'$
  is a constant which is independent of~$L$.
\end{theorem}
\begin{proof}
  Let us introduce a parameter $\lambda$ and then expand in powers of $\lambda$
  to obtain
  \begin{displaymath}
    \left|\det\left(1+{\lambda\over L}\Delta\right) \right|
      = \left| \sum_{k=0}^L \left(\lambda\over L\right)^k
	\sum_{i_1>i_2>\cdots>i_k} \det\left(
	\begin{array}{ccc}
	  \Delta_{i_1i_1} & \ldots & \Delta_{i_1i_k} \\
	  \vdots & \ddots & \vdots \\
	  \Delta_{i_ki_1} & \ldots & \Delta_{i_ki_k} \\
	\end{array}
      \right) \right|,
  \end{displaymath}
  using Hadamard's bound
  \begin{displaymath}
    \leq\, \sum_{k=0}^L \left(\lambda\over L\right)^k
      \sum_{i_1>i_2>\cdots>i_k}(C\sqrt k)^k
    \,=\, \sum_{k=0}^L \left(\lambda\over L\right)^k {L\choose k} (C\sqrt k)^k
    \,\leq\, \sum_{k=0}^\infty {(\lambda C\sqrt k)^k\over k!}
    \,=\, C'(\lambda)
  \end{displaymath}
  where the ratio test may be applied to show that the series converges for all
  $\lambda$. The desired result follows upon setting $\lambda=1$.
\end{proof}

\begin{definition}
  The {\em minor} $\Delta^{(ij)}$ of a matrix $\Delta$ is defined by
  \begin{displaymath}
    \Delta^{(ij)}_{kl} \equiv \left\{
      \begin{array}{cl}
	1 & \mbox{if $k=i$ and $j=l$} \\
	\Delta_{kl} & \mbox{if $k\neq i$ and $j\neq l$} \\
	0 & \mbox{otherwise} \\
      \end{array}
      \right.
  \end{displaymath}
\end{definition}

\begin{corollary}[Fredholm] \label{corollary:fredholm}
  Let $\Delta$ be a matrix as in Theorem~\ref{theorem:fredholm}, then
  \begin{displaymath}
    \left|\det\left(\identity + {1\over L}\Delta^{(ij)}\right)\right|
      \leq \left\{
      \begin{array}{cl}
	C' & \mbox{if $i=j$} \\
	C'/L & \mbox{if $i\neq j$} \\
      \end{array}
      \right.
  \end{displaymath}
  where $C'$ is an ($L$-independent) constant.
\end{corollary}
\begin{proof}
  If $i=j$ then the matrix $\identity+{1\over L}\Delta^{(ij)}$ is of the form to
  which Theorem~\ref{theorem:fredholm} is applicable. If $i\neq j$ then
  introducing the parameter $\lambda$ and expanding in powers of $\lambda$ as
  before we obtain
  \begin{displaymath}
    \left|\det\left(1+{\lambda\over L}\Delta^{(ij)}\right) \right|
      = \left| \sum_{k=0}^{L-2} \left(\lambda\over L\right)^{k+1}
	\sum_{{\scriptstyle i_1>i_2>\cdots>i_k \atop
	  \scriptstyle i_r>i>i_{r+1}} \atop
	  \scriptstyle i_s>j>i_{s+1}} (-1)^{r+s} \det\left(
	\begin{array}{cccc}
	  \Delta_{ji} & \Delta_{ji_1} & \ldots & \Delta_{ji_k} \\
	  \Delta_{i_1i} & \Delta_{i_1i_1} & \ldots & \Delta_{i_1i_k} \\
	  \vdots & \vdots & \ddots & \vdots \\
	  \Delta_{i_ki} & \Delta_{i_ki_1} & \ldots & \Delta_{i_ki_k} \\
	\end{array}
      \right) \right|,
  \end{displaymath}
  using Hadamard's bound
  \begin{eqnarray*}
    &\leq& \sum_{k=0}^{L-2} \left(\lambda\over L\right)^{k+1}
	\sum_{{\scriptstyle i_1>i_2>\cdots>i_k \atop
	  \scriptstyle i_r>i>i_{r+1}} \atop
	  \scriptstyle i_s>j>i_{s+1}} (C\sqrt{k+1})^{k+1}
    \,=\, \sum_{k=0}^{L-2} \left(\lambda\over L\right)^{k+1} {L-2\choose k}
      (C\sqrt{k+1})^{k+1} \\
    &\leq& {1\over L} \sum_{k=0}^\infty {(\lambda C\sqrt{k+1})^{k+1}\over k!}
    \,=\, {C'(\lambda)\over L}
  \end{eqnarray*}
  where the ratio test may be applied to show that the series converges for all
  $\lambda$. The desired result follows upon setting $\lambda=1$.
\end{proof}

\begin{definition}
  The {\em adjoint} $\Adj M$ of a matrix $M$ is defined to be the transpose of
  the matrix whose elements are the determinants of the corresponding minors,
  that is $(\Adj M)_{ij} \equiv \det M^{(ji)}$.
\end{definition}

\begin{theorem} \label{theorem:inverse}
  Let $\Delta:l^2_L(\C)\to l^2_L(\C)$ be a matrix all of whose elements are
  bounded by some constant $|\Delta_{ij}|\leq C$, and $\det\left(\identity+
  {1\over L}\Delta\right)\geq \alpha>0$ is bounded below by an ($L$-independent)
  constant, then $\left(\identity+{1\over L}\Delta\right)^{-1} = \identity +
  {1\over L}G$, where all the matrix elements of $G$ are bounded by an
  ($L$-independent) constant, $|G_{ij}|\leq C''$.
\end{theorem}
\begin{proof}
  From the algebraic identity (Cramer's rule)
  \begin{displaymath}
    \left(\identity + {1\over L}\Delta\right)
      \Adj\left(\identity + {1\over L}\Delta\right)
      = \det\left(\identity + {1\over L}\Delta\right) \identity
      \equiv D \identity
  \end{displaymath}
  and the fact that $\left[ \Adj\left(\identity + {1\over L}\Delta\right)
  \right]_{ij} = C_{ij}/L$ for $i\neq j$ and $\left[ \Adj\left(\identity +
  {1\over L}\Delta\right) \right]_{jj} = C_{jj}$, which follow from
  Corollary~\ref{corollary:fredholm}, we have
  \begin{displaymath}
    \left(1 + {1\over L}\Delta_{jj}\right) C_{jj}
      + \sum_{i\neq j} {1\over L^2} \Delta_{ji} C_{ij} = D.
  \end{displaymath}
  From this we find that $C_{jj} = D + O(1/L) \implies C_{jj}/D = 1 + O(1/L)$,
  where we have made use of the fact that $D$ is bounded below by the constant
  $\alpha$. The desired result then follows from substituting this relation into
  Cramer's rule above.
\end{proof}

\begin{lemma} \label{lemma:inf-vol-lim}
  If the matrix elements $\Delta_{ij}$ referred to above are taken to be the
  values $\Delta\left({2\pi i\over L},{2\pi j\over L}\right)$ of a Lebesgue
  integrable function then we can define the infinite volume limit of the
  Fredholm determinant, and if this limit is non-zero then there must exist a
  non-zero constant $\alpha$ which bounds it from below.
\end{lemma}

\begin{definition}
  Let $\alpha$ be a subset of the integers $\{1,\ldots,d\}$, and
  $\delta^{(\alpha)}_{pq}\equiv\prod_{\mu\not\in\alpha}\delta_{p_\mu q_\mu}$. A
  matrix $A$ has support on dimensions $\alpha$ if $A^{(\alpha)}_{pq} =
  \delta^{(\alpha)}_{pq} A_{pq}$, with $p$ and $q$ being a $d$-tuple of indices,
  $p=(p_1,\ldots,p_d)$.
\end{definition}
Notice that according to our definition $A^{(\emptyset)}$ is a diagonal matrix.
\begin{corollary} \label{corollary:ivan}
  There is a permutation of the indices $p$ for each matrix $A^{(\alpha)}$ such
  that it is block diagonal with $\#\alpha\times\#\alpha$~blocks. Each block
  satisfies the assumptions of Theorem~\ref{theorem:inverse}, so the inverse may
  be expressed in block diagonal form where each block is of the form specified
  in Theorem~\ref{theorem:inverse}.
\end{corollary}

\begin{lemma} \label{lemma:multiplication}
  If $A^{(\alpha)}$ has support on dimensions $\alpha$ and $B^{(\beta)}$ has
  support on dimensions $\beta$, and all the elements of are bounded by an
  ($L$-independent) constant, $|A^{(\alpha)}_{pq}|\leq C$ and
  $|B^{(\beta)}_{pq}|\leq C$, then
  \begin{displaymath}
    \sum_q L^{-\#\alpha} A^{(\alpha)}_{pq} L^{-\#\beta} B^{(\beta)}_{qr}
      = L^{-\#\alpha\union\beta} W^{(\alpha\union\beta)}_{pr}
  \end{displaymath}
  where $W^{(\alpha\union\beta)}$ has support on dimensions $\alpha\union\beta$
  and $|W^{(\alpha\union\beta)}_{pr}|\leq C'$. $\#\alpha$ is the cardinality of
  the set $\alpha$.
\end{lemma}
\begin{proof}
  Consider the matrix product
  \begin{displaymath}
    \sum_q  A^{(\alpha)}_{pq} B^{(\beta)}_{qr}
      = \sum_q A_{pq} B_{qr} \prod_{\mu\not\in\alpha\atop\nu\not\in\beta}
	\delta_{p_\mu q_\mu} \delta_{q_\nu r_\nu}.
  \end{displaymath}
  The product of Kronecker tensors may be split into four disjoint classes,
  $\overline{\alpha\union\beta}$, $\beta-\alpha$, $\alpha-\beta$, and
  $\alpha\intersection\beta$:
  \begin{displaymath}
    \left(A^{(\alpha)} B^{(\beta)}\right)_{pr}
      = \sum_q A_{pq} B{qr} \left(\prod_{\mu\not\in\alpha\union\beta}
	\delta_{p_\mu q_\mu} \delta_{q_\mu r_\mu}\right)
	\left(\prod_{\mu\in\beta-\alpha} \delta_{p_\mu q_\mu}\right)
	\left(\prod_{\mu\in\alpha-\beta} \delta_{q_\mu r_\mu}\right)
	\left(\prod_{\mu\in\alpha\intersection\beta} 1\right).
  \end{displaymath}
  It is immediately apparent that only for $\mu$ in the last set of indices
  ($\alpha\intersection\beta$) does the sum over $q_\mu$ have more than one
  non-vanishing term, and that for $\mu$ in the first set of indices ($\overline
  {\alpha\union\beta}$) the expression vanishes unless $p_\mu=r_\mu$. We thus
  have
  \begin{displaymath}
    \left(A^{(\alpha)} B^{(\beta)}\right)_{pr}
      = \prod_{\mu\not\in\alpha\union\beta} \delta_{p_\mu r_\mu}
	L^{\#\alpha\intersection\beta} W_{pr},
  \end{displaymath}
  with $|W_{pr}|\leq C'$. The result now follows from the observation that
  $\#\alpha\union\beta = \#\alpha + \#\beta - \#\alpha\intersection\beta$.
\end{proof}

\begin{theorem} \label{theorem:main}
  Let $\Delta^{(\alpha)}$ be a matrix with support on dimensions $\alpha$ all of
  whose elements are bounded by an ($L$-independent) constant, and furthermore
  \begin{displaymath}
    \lim_{L\to\infty}\det\left(\identity+ \sum_{\alpha\neq\emptyset}
      L^{-\#\alpha} \Delta^{(\alpha)}\right)\neq0;
  \end{displaymath}
  then
  \begin{displaymath}
    \left(\identity
	+ \sum_{\alpha\neq\emptyset} L^{-\#\alpha} \Delta^{(\alpha)}\right)^{-1}
      = \identity + \sum_{\alpha\neq\emptyset} L^{-\#\alpha} G^{(\alpha)}
  \end{displaymath}
  where $G^{(\alpha)}$ its has support on dimensions $\alpha$ and all its
  elements bounded by a constant.
\end{theorem}
\begin{proof}
  We may order the subsets of $\{1,\ldots,d\}$ lexicographically, that is
  $\alpha<\beta$ iff $\#\alpha<\#\beta$ or $\#\alpha=\#\beta$ and $\min_{i\in
  \alpha-\beta}i < \min_{j\in\beta-\alpha}j$. Let $\gamma$ be the smallest set
  for which $\Delta^{(\gamma)}\neq0$. Then
  \begin{displaymath}
    \identity + \sum_{\alpha\neq\emptyset} L^{-\#\alpha} \Delta^{(\alpha)}
      = \left(\identity + L^{-\#\gamma} \Delta^{(\gamma)}\right)
	\left(\identity + \left(\identity + L^{-\#\gamma}
	    \Delta^{(\gamma)}\right)^{-1}
	  \sum_{\alpha>\gamma} L^{-\#\alpha} \Delta^{(\alpha)}\right).
  \end{displaymath}
  The infinite volume limit of the determinant of this quantity is non zero,
  therefore the infinite volume limit of the determinant of each factor must be
  non zero. From Corollary~\ref{corollary:ivan} we have that $(\identity +
  L^{-\#\gamma} \Delta^{(\gamma)})^{-1} = \identity + L^{-\#\gamma}
  G^{(\gamma)}$ with $|G_{pq}|\leq C'$, hence
  \begin{eqnarray*}
    \left(\identity + \sum_{\alpha\neq\emptyset} L^{-\#\alpha}
	\Delta^{(\alpha)}\right)^{-1}
      &=& \left(\identity + \left(\identity + L^{-\#\gamma} G^{(\gamma)}\right)
	  \sum_{\alpha>\gamma} L^{-\#\alpha} \Delta^{(\alpha)}\right)^{-1}
	  \left(\identity + L^{-\#\gamma} G^{(\gamma)}\right) \\
      &=& \left(\identity + \sum_{\alpha>\gamma}
	L^{-\#\alpha} {\Delta'}^{(\alpha)}\right)^{-1}
	  \left(\identity + L^{-\#\gamma} G^{(\gamma)}\right) \\
      &=& \identity + \sum_{\alpha\neq\emptyset}
	L^{-\#\alpha} {G'}^{(\alpha)}
  \end{eqnarray*}
  where we have used Lemmas~\ref{lemma:inf-vol-lim}
  and~\ref{lemma:multiplication}, and the fact that
  $\alpha>\gamma\implies\alpha\union\gamma>\gamma$ in the penultimate line. The
  final line follows by induction on $\gamma$.
\end{proof}

One application of these results uses the matrices $N$, $O$, $N^D$, $O^D$, and
$R$ introduced in the body of the paper, and the matrix $\Delta$ defined as
$\Delta\equiv(\identity-N^D)^{-1}R$.
\begin{corollary} \label{corollary:M-form}
  $M$ has the form given in equation~(\ref{eq:x7}).
\end{corollary}
\begin{proof}
  Observe that
  \begin{displaymath}
    \identity-N = \left(\identity - N^D + {1\over L}R\right)
      = (\identity-N^D)
	\left(\identity	+ {1\over L}(\identity-N^D)^{-1} R\right)
      = (\identity-N^D) \left(\identity + {1\over L}\Delta\right);
  \end{displaymath}
  since there is a basis in which $N$ is a strictly triangular matrix it follows
  that $\det(\identity-N)=1\,(\forall L)$, and thus we may conclude that
  $\lim_{L\to\infty}\det\left(\identity + {1\over L}\Delta\right)\neq0$.
  Therefore $\Delta$ satisfies the assumptions of Theorem~\ref{theorem:main},
  so $\left(\identity - {1\over L}\Delta\right)^{-1}$ has the desired form.
  From equation~(\ref{eq:y1}) we have
  \begin{displaymath}
    M = \identity - (\identity-N)^{-1} (\identity - N^D - O^D)
      = \identity - \left(\identity + {1\over L}\Delta\right)^{-1}
	(\identity-N^D)^{-1} (\identity - N^D - O^D),
  \end{displaymath}
  so we may apply Lemma~\ref{lemma:multiplication} to show that $M$ also has the
  desired form.
\end{proof}

If the infinite volume Markov process has a unique fixed point then from
equation~(\ref{eq:av-quad-update}) we obtain upon averaging over this fixed
point distribution of~$\phi$
\begin{displaymath}
  (\identity - M^Q) \langle\phi^Q\rangle_{\phi,\pi} = P^Q,
\end{displaymath}
and the uniqueness of $\langle\phi^Q\rangle_{\phi,\pi}$ implies that
$\lim_{L\to\infty}\det\left(\identity - M^*\otimes M\right)\neq0$. As a
consequence of Corollary~\ref{corollary:M-form} we may write
\begin{displaymath}
  \identity - M^*\otimes M = \left(\identity - {M^D}^*\otimes M^D\right)
    \left(\identity + \sum_\alpha L^{-\#\alpha} A^{(\alpha)}\right),
\end{displaymath}
where the matrix elements of $A^{(\alpha)}$ are bounded by an $L$-independent
constant, and thus from Theorem~\ref{theorem:main} and
Lemma~\ref{lemma:multiplication} we may conclude that a result analogous to
Corollary~\ref{corollary:M-form} applies to quadratic operators.

\bibliographystyle{utphys}
\bibliography{adk,lattice-bibliography}
\end{document}
\bye